\documentclass[journal,comsoc]{IEEEtran}
\usepackage[font=small]{caption}
\usepackage[T1]{fontenc}% optional T1 font encoding
\usepackage{cite}
\usepackage{amsmath,amssymb,amsfonts}
\usepackage{algorithmic}
\usepackage{graphicx}
\usepackage{textcomp}
\usepackage{xcolor}
\usepackage{glossaries}
\usepackage{hyperref}
\usepackage{cleveref}
\usepackage{bm}
\usepackage[algoruled, vlined, linesnumbered]{algorithm2e}
\usepackage{algorithmic}
\usepackage{subfig}

\usepackage[cmintegrals]{newtxmath}

\DeclareMathOperator*{\argmax}{arg\,max}

\newcommand{\xb}{\bm{x}}
\newcommand{\db}{\bm{\delta}}
\newcommand{\ob}{\mathbf{o}}

\newcommand{\ind}{1{\hskip -2.5 pt} \mathrm{I}}
\newacronym{5gnr}{5GNR}{5G New Radio}
\newacronym{ack}{ACK}{acknowledgment}
\newacronym{api}{API}{application programming interface}
\newacronym{bler}{BLER}{block error rate}
\newacronym{bo}{BO}{Bayesian optimization}
\newacronym{bogp}{BOGP}{Bayesian optimisation with Gaussian processes}
\newacronym{bs}{BS}{base station}
\newacronym{cd}{CD}{coordinate descent}
\newacronym{clpc}{CLPC}{closed-loop power control}
\newacronym{csma}{CSMA}{carrier-sense multiple access}
\newacronym{cson}{C-SON}{centralised self organizing networks}
\newacronym{dl}{DL}{downlink}
\newacronym{fdm}{FDM}{frequency division multiplexing}
\newacronym{fdma}{FDMA}{frequency division multiple access}
\newglossaryentry{gp}
{
  name={GP},
  description={Gaussian process},
  first={\glsentrydesc{gp} (\glsentrytext{GP})},
  plural={GPs},
  firstplural={\glsentrydesc{gp}es (\glsentryplural{gp})}
}
\newacronym{gss}{GSS}{golden-section search}
\newacronym{kpi}{KPI}{key performance indicator}
\newacronym{l2c}{L2C}{learning-to-communicate}
\newacronym{lbt}{LBT}{listen before talk}
\newacronym{lte}{LTE}{Long Term Evolution}
\newacronym{los}{LoS}{line of sight}
\newacronym{mac}{MAC}{medium access control}
\newacronym{madrl}{MADRL}{multi-agent deep reinforcement learning}
\newacronym{marl}{MARL}{multi-agent reinforcement learning}
\newacronym{ml}{ML}{machine learning}
\newacronym{mno}{MNO}{mobile network operator}
\newacronym{mumimo}{MU-MIMO}{multi-user multiple-input multiple-output}
\newacronym{olpc}{OLPC}{open-loop power control}
\newacronym{pdu}{PDU}{protocol data unit}
\newacronym{pm}{PM}{performance management}
\newacronym{phy}{PHY}{physical layer}
\newacronym{pomdp}{POMDP}{partially observable Markov decision process}
\newacronym{prb}{PRB}{physical resource block}
\newacronym{pusch}{PUSCH}{physical uplink shared channel}
\newacronym{rl}{RL}{reinforcement learning}
\newacronym{rrm}{RRM}{radio resource management}
\newacronym{sdu}{SDU}{service data unit}
\newacronym{sg}{SG}{scheduling grant}
\newacronym{sinr}{SINR}{signal-to-noise-plus-interference ratio}
\newacronym{son}{SON}{self organizing networks}
\newacronym{sr}{SR}{scheduling request}
\newacronym{tdm}{TDM}{time division multiplexing}
\newacronym{tdma}{TDMA}{time division multiple access}
\newacronym{ue}{UE}{user equipment}
\newacronym{ul}{UL}{uplink}
\newacronym{ulpc}{ULPC}{uplink power control}
\newacronym{umi}{UMi}{urban micro}
\newacronym{urllc}{URLLC}{ultra reliable low latency communications}

\hypersetup{
pdftitle={Bayesian Optimization for Radio Resource Management: Open Loop Power Control},
pdfsubject={Machine Learning for communications},
pdfauthor={Lorenzo Maggi, Alvaro Valcarce, Jakob Hoydis},
pdfkeywords={}
}

\begin{document}
\title{Bayesian Optimization for Radio Resource Management: Open Loop Power Control}

\author{Lorenzo~Maggi,
			Alvaro~Valcarce,~\IEEEmembership{Senior~Member,~IEEE,}
        and~Jakob~Hoydis,~\IEEEmembership{Senior~Member,~IEEE}
				\thanks{The authors are with Nokia Bell-Labs France, Route de Villejust, 91620 Nozay, France (Email: \{lorenzo.maggi, alvaro.valcarce\_rial, jakob.hoydis\}@nokia-bell-labs.com).}% 
}

\maketitle

\begin{abstract}
We provide the reader with an accessible yet rigorous introduction to \gls{bogp} for the purpose of solving a wide variety of \gls{rrm} problems. We believe that \gls{bogp} is a powerful tool that has been somewhat overlooked in \gls{rrm} research, although it elegantly addresses pressing requirements for fast convergence, safe exploration, and interpretability. \Gls{bogp} also provides a natural way to exploit prior knowledge during optimization. After explaining the nuts and bolts of \gls{bogp}, we delve into more advanced topics, such as the choice of the acquisition function and the optimization of dynamic performance functions. Finally, we put the theory into practice for the \gls{rrm} problem of uplink \gls{olpc} in 5G cellular networks, for which \gls{bogp} is able to converge to almost optimal solutions in tens of iterations without significant performance drops during exploration.
\end{abstract}


\glsresetall

\section{Introduction}
The current generations of cellular networks (4G \& 5G) specify a plethora of options and parameters to adapt them to different frequency bands and a wide range of deployment scenarios. However, how to choose these parameters optimally when deploying and operating a commercial network is not specified. This problem is generally referred to as \gls{rrm}, whose complexity has grown from generation to generation. Such trend is likely to continue for 6G which is expected to serve an even larger diversity of spectrum, deployment options, and use cases \cite{viswanathan2020}. In order to deal with this growing complexity, \gls{ml} methods are seen as a key technology and, consequently, a large number of vision and survey papers have been published during the last few years on this topic, e.g., \cite{8844682, letaief2019roadmap}.     
Yet, several challenges to the use of \gls{ml} for \gls{rrm} in commercial networks are frequently pointed out \cite{8994961, 2020katotenchallenges}: convergence speed, safe exploration and interpretability of solutions. 

The first challenge relates to the time or number of trials it takes to find a good set of parameters. This is not an issue in simulation-based studies, but of utmost importance for realistic deployments. While \gls{rl} has shown a great potential to solve \gls{rrm} problems in simulations by running  algorithms for thousands of iterations (e.g., \cite{calabrese2018learningrrm}), it often suffers from poor convergence speed which renders its use questionable for many real-world applications, where one is often limited to a few tens of iterations.   
The second challenge is to avoid that an \gls{ml} algorithm chooses \gls{rrm} parameters that lead to unacceptably low performance levels, even if only chosen temporarily as an intermediate step towards better settings. Again, while this is not an issue in simulations, it is another key requirement of \glspl{mno} which will ultimately decide if a solution is adopted in practice or not. 
The last challenge relates to understanding why an \gls{ml} algorithm has chosen a specific parameter setting. This is not necessarily an issue when the first two points are well addressed, but certainly of help for debugging purposes and to increase trust in \gls{ml} methods in general. 

Although these challenges are widely known and discussed, it is surprising that little research is done to explore the use of \gls{ml} methods which are better suited to such practical constraints. In our experience, \gls{bogp}\footnote{Throughout the paper, we employ indifferently the acronyms BOGP and BO (Bayesian optimization).} is a very powerful tool, which is well adapted to a wide range of \gls{rrm} parameter tuning tasks, and that addresses the above challenges to a large extent.
As shown recently in \cite{Dreifuerst2020} for the task of transmit power and antenna tilt optimization, \gls{bogp} can achieve convergence within two orders of magnitude fewer iterations than a modern \gls{rl}-based solution. A remarkable additional advantage of \gls{bogp} is that it allows to naturally embed prior knowledge from experts, simulations, or real-world deployments in the optimization process.
The basics of \gls{bogp} are already described in \cite{Shahriari2016}. In addition, our paper addresses the difficulties that \gls{bogp} practitioners will encounter when applying it to \gls{rrm} problems.
This includes topics such as prior reuse for faster convergence, time evolving objective functions and the choice of a suitable acquisition function.

The goal of this paper is hence two-fold. First, we aim to provide an intuitive yet detailed user guide to \gls{bo} with \glspl{gp}. Second, we claim that \gls{bo} is a useful but yet largely untapped algorithmic tool for \gls{rrm} in general, and we demonstrate it via a specific use-case, namely that of \gls{olpc} in the uplink.
In Section~\ref{sec:bogp}, we aim to give the reader the ability to understand whether the \gls{rrm} problem at hand can/should be solved via BO. Then, if the answer is positive, we invite the reader to embark on a more technically involved journey through the fundamentals of the theory of \glspl{gp} and \gls{bo} in Section~\ref{sec:GPtheory}. Section~\ref{sec:BOdetails} describes aspects of \gls{bogp} for \gls{rrm} problems, while Section~\ref{sec:ULPC} focuses explicitly on \gls{olpc}. The paper is concluded in Section~\ref{sec:conclusions}.

\section{Is Bayesian Optimization suited for your radio resource management problem?}\label{sec:bogp}

It is convenient at this early stage to outline the general \gls{rrm}{} framework that we will refer to throughout the paper, and that we will instantiate via uplink power control.\footnote{Although our focus is on \gls{rrm} problems, most of the following discussion is also valid for the broader class of resource management problems.}

\textbf{\gls{rrm} problem:} There exists a set of $k$ parameters, described by the vector $\xb\in \mathbb R^{k}$, that a radio system hinges upon and that a controller wants to optimize. We call $\mathcal X\subset \mathbb{R}^k$ the set of available parameter configurations. We remark that $\mathcal X$ can be a discrete set. The performance of the system, when the parameters take on the value $\xb$, is described by a single real value $f(\xb)\in \mathbb R$. The function $f(.)$ is \emph{unknown} to the controller and is possibly \emph{noisy}. We call $\widetilde{f}(.)$ its realizations, and suppose $\mathbb E[\widetilde{f}(.)]=f(.)$. 
The controller is \emph{allowed} to sequentially test different parameter values $\xb(1),\xb(2),\dots$, and it is \emph{able} to observe the corresponding performance realizations $\widetilde{f}(\xb(1)),\widetilde{f}(\xb(2)),\dots$. 
The general \emph{goal} is to ensure that performance is good at any point in time, while possibly converging to the optimal configuration $\xb^*=\argmax_{\xb} f(\xb)$ and, most importantly, without incurring sudden drops in performance.\\

A more difficult variant of this problem arises when the function $f(.)$ varies over time: 

\textbf{Dynamic \gls{rrm} problem:} Assume that at any time $t$, there exists a \emph{hidden} system ``state'' $\theta(t)$ that varies over time \emph{independently} of the deployed \gls{rrm}{} configuration. Then, the system performance is a function $f(\xb,\theta(t))$ of the hidden state. When the controller deploys at time $t$ the configuration $\xb$, it only observes $f(\xb,\theta(t))$ but it is agnostic to $\theta(t)$.\\

We deliberately decide not to make the concept of ``state'' $\theta$ explicit at this stage. One can think of $\theta$ as the collection of random environmental variables that evolve \emph{exogenously} (e.g., user geographical distribution, cell load, etc.), that affect the system performance, and that are hard to identify \emph{a priori}. We will come back to the dynamic \gls{rrm} problem in Section~\ref{sec:timevarying}, where our approach does not prescribe to infer $\theta(t)$ itself, but rather to learn its speed of variation---and if any, its periodicity---and track the time-varying maximum of the function $f(.,\theta(t))$. 

\subsection{What \underline{must} hold for BO to be a suitable choice} \label{sec:rules_of_thumb}

In order for BO to be considered as a viable solution for the general \gls{rrm} problem defined above, the latter must fulfill two main conditions. We argue that the first condition is the most binding one.

\textbf{C1}. \textbf{Number of control parameters:} The theory of \gls{bo} is general enough to account for any (finite) number $k$ of parameters to be jointly optimized. In practice, however, $k$ must be ``small'', i.e., of the order of 20 as a rule of thumb \cite{frazier2018tutorial}. As explained in Section \ref{sec:BOdetails}, the motivation is that a non-convex problem in $\mathbb{R}^k$ has to be solved for choosing the next \gls{rrm} configuration to be deployed in the system.
	
\textbf{C2}. \textbf{Smoothness of performance function:} The performance function $f$ must not have discontinuities with respect to the control parameters $\xb$. Indeed, BO strongly leverages the assumption that the function $f$ is sufficiently smooth to be able to infer it \emph{everywhere} (i.e., for all possible values of $\xb$) by just sampling it at a (possibly) small number of points. \\

Although the theory falls apart when the performance function $f$ presents discontinuities (C2), \gls{bo} can still be used in practice, as long as the number of discontinuities is limited \cite{cornford1998adding}. However, if the number of parameters is high (C1), we argue that \gls{bo} is not a suitable tool for the \gls{rrm} problem at hand.

\subsection{What \underline{should} hold for \gls{bo} to be a suitable choice} \label{sec:should}

We claim that there exists one additional---yet hard to quantify \emph{a priori}---condition that should be met for \gls{bo} to be successfully applicable to \gls{rrm} problems.

\textbf{C3}. \textbf{Update frequency:} The control parameters $\xb$ are to be modified at a relatively ``low'' frequency. It is hard to provide a rule of thumb for this, as it is dependent on the application. In general, it is a natural consequence of two factors:
\begin{itemize}
	\item[i)] At iteration $n$, once the parameter configuration $\xb(n)$ is deployed in the system, the metric $\widetilde{f}(\xb(n))$ is collected. Note that $\widetilde{f}(\xb)$ is meant as a sample average of the system performance computed while $\xb$ is deployed in the system. The less noisy $\widetilde{f}(.)$ is, the better the \gls{gp} manages to infer the function $f(.)$, and the quicker \gls{bo} converges to the optimal configuration. Hence, updating $\xb$ at a lower frequency allows one to observe and average out the performance metric across a longer time span, and eventually to improve the GP inference.
	\item[ii)] At each iteration, deciding which point to try next is a task whose computational effort increases with the number of collected samples, and it generally requires to solve a non-convex problem. As a rule of thumb, for a reasonable range of applications with at most 100 different sampled configurations and 10 variables to optimize, one iteration may require a few seconds.\footnote{This depends of course on the available compute power.}
\end{itemize}

\subsection{The added value}
A prominent feature of \gls{bo} consists in providing a principled approach to embed any prior information that the controller \emph{may} possess on the shape of the performance function $f(.)$. Such prior can be expressed with two auxiliary functions.

\textbf{C4}. \textbf{Mean prior:} A prior function $m(.)$ on the mean prior should reflect a ``belief'' on the unknown function $f$ at a point $\xb$. Intuitively, one would want $m(\xb)\approx f(\xb)$; by default, if no prior information is available, then one sets $m(\xb)=0$ for all possible values of $\xb$.

\textbf{C5}. \textbf{Covariance kernel:} A prior function $C(.,.)$ on the covariance of $f$ should describe how much we expect the system performance to differ when the configurations $\xb$ and $\xb'$ are deployed, respectively. In other words, one wishes that $C(\xb,\xb')\approx \mathrm{Cov}\big[f(\xb),f(\xb')\big]$. In practice, one should define $C(\xb,\xb')$ through \emph{kernel} functions that only depend on the ``distance'' between $\xb$ and $\xb'$ (see Section~\ref{sec:GPtheory}).\\

Both the mean and the kernel are to be designed, e.g., by exploiting historical data or via expert domain knowledge.
We remark that their specification is \emph{not}  necessary for a correct functioning of BO---in principle, they can be set to any default value---but it is still recommended if one wants to boost the convergence properties of the BO algorithm. Indeed, as we illustrate in Section\ref{sec:results} via simulation, possessing a good prior \emph{mean} function allows BO to avoid random exploration and sudden drops in system performance. On the other hand, as showed in Section \ref{sec:hyparam_tuning}, the covariance kernel typically depends on hyperparameters that are optimized at run-time, as new observations are collected, which makes the prior choice for the covariance function $C$ less of a delicate task.

\subsection{The benefits of \gls{bo}} \label{sec:benefitsBO}
We now assume that conditions C1-C2 hold, and hopefully condition C3 is also verified. Then, we claim that \gls{bo} can be successfully applied to solve the general \gls{rrm} problem outlined at the beginning of this section. Moreover, the following properties can be appreciated which are particularly important for solving real-life problems:

\textbf{P1}. \textbf{Quick convergence:} \gls{bo} converges to a near optimal configuration in few iteration steps.

\textbf{P2}. \textbf{Safe exploration:} \gls{bo} avoids sudden drops in the experienced system performance since it searches across ``safe'' configuration parameter regions.

\textbf{P3}. \textbf{Risk-vs.-return for untested configurations:} \gls{bo} is able to quantify the risk-vs.-return trade-off related to choosing \emph{any} configuration parameter $\xb$ (especially those which have \emph{never} been deployed in the past).

We argue that the properties above are key in realistic \gls{rrm}{} applications. They are also classic pain points for other optimization techniques which are more widely known in the \gls{rrm}{} community, as illustrated next.

\subsection{Available alternatives to \gls{bo}}

We now discuss alternatives to \gls{bo} that are available in the literature and that could be employed to tackle our general \gls{rrm}{} problem. We first observe that the controller is generally agnostic to the derivative of the performance function $f(.)$---in fact, $f(.)$ itself is supposed to be unknown---and can only observe its (noisy) samples at given configuration points. For this reason, the whole class of gradient descent optimization algorithms is not a viable option. Still, one could argue that the gradient could be approximated by finite differences. However, this clearly slows down convergence since $\mathcal O(k)$ samples are used only for estimating the gradient, and moreover the observation noise can wildly corrupt the gradient estimate. 

Turning to machine-learning (ML) based techniques that are widely known in the \gls{rrm}{} community, multi-armed bandits (MAB) \cite{cesa2006prediction} could be in principle applied to our problem. E.g., one could discretize the parameter configuration space $\mathcal X$ and consider each available configuration as an independent arm. This approach is agnostic to the derivative of $f(.)$, nonetheless it presents a major drawback in that MAB requires to sample each and every single arm at least once to accumulate sufficient statistics, i.e., it is not able to infer the quality of an \gls{rrm} configuration that has never been tested. This is clearly unacceptable for real-world problems where convergence speed is key and properties P1-P2 are highly appreciated.

Another popular ML technique that is nowadays widely applied in the \gls{rrm}{} space is Reinforcement Learning (\gls{rl}) \cite{sutton2018reinforcement}. Yet, \gls{rl} is particularly suited for time-varying reward (i.e., performance) functions where there exists a ``state'' that evolves according to exogenous \emph{and} endogenous factors, i.e., it depends on the \gls{rrm} control strategy of the controller as well. However, our primary focus is on static reward functions. Moreover, the time-varying \gls{rrm} problem that we defined considers the evolution of the state $\theta(t)$ as independent of the \gls{rrm} strategy, hence RL would be an overkill for our problem since there is no long-term strategic planning involved. Last but not least, it is widely known that RL generally suffers from serious convergence issues, for which it is inevitable to incur harsh performance drops during an---at least---initial exploration phase. Hence, this hinders us to achieve the qualitative convergence properties P1-P2. We remark that the RL community is  of course aware of this issue and currently deploys important efforts to tackle it \cite{garcia2015comprehensive}.

Let us turn our attention towards the realm of the so-called \emph{derivative-free} optimization methods \cite{conn2009introduction}, which are completely agnostic to the function to be optimized. They can be classified into i) \emph{direct search} methods, deciding the next point to sample only based on the value of the function at previously sampled points, without any explicit or implicit derivative approximation or function model building; ii) \emph{trust-region} methods, defining a region around the current best solution, in which a (usually quadratic) model is built to approximate the unknown objective function and to decide the next point; iii) \emph{surrogate} models, constructing a surrogate model of the unknown function at \emph{all} points---hence not necessarily close to the current best solution. 

Surrogate models are particularly suited when one can only afford to test a very small set of points before landing on a reasonably good solution. This typically occurs when the objective function is expensive to evaluate \cite{audet2017derivative}, e.g., for real-world applications like probe drilling for oil at given geographic coordinates, evaluating the effectiveness of a drug candidate, or testing the effectiveness of deep neural network hyperparameters. 
In our case, testing a new \gls{rrm} configuration parameter is not expensive \textit{per se}, but the controller still wants to greatly limit the exploration of new parameters and quickly converge to a reasonably good solution, leading to an equivalent concept of ``expensive'' objective function. \gls{bo} is arguably the most prominent example of surrogate models, where the surrogate function is usually a \gls{gp}. For this reason, we will next recap the theoretical foundations of \glspl{gp}.

\section{Modeling the performance function as a Gaussian process (\gls{gp})} \label{sec:GPtheory}

In a nutshell, a \gls{gp} is a collection of random variables, any finite number of which have a Gaussian distribution \cite{williams2006gaussian}. In this paper, we propose to model the system performance function $f(.)$ as a \gls{gp}, since \glspl{gp} are a powerful inference engine allowing to predict the value of the function $f(.)$ for \gls{rrm} configurations that have never been deployed in the system.
\glspl{gp} are indeed the main sub-routine of \gls{bo} \cite{theodoridis2015machine}, whose task is in turn to suggest to the controller the next configuration parameter to deploy in the network. This is achieved on the basis of the \gls{gp} model that has been learned via the record of system performance observations resulting from the previously deployed \gls{rrm} configurations.

Clearly, in practice the system performance function $f$ may follow a distribution which is not Gaussian. However, \glspl{gp} have been shown to provide an excellent trade-off between modeling accuracy and inference complexity. For this reason, we will only focus on \glspl{gp} as the main BO inference engine. Yet, it is worth mentioning that Student-$t$ processes are a valid alternative, as studied, e.g., in \cite{shah2014student}.

Before delving into the details of \glspl{gp}, we deem it convenient to first review some basics on multivariate Gaussian distributions.

\subsection{Multivariate Gaussian distribution} \label{sec:gauss_intro} 

A random variable $Y$ with $d$ components is a Gaussian variable if its probability distribution $p(.)$ can be written as
\begin{equation}
p(\mathbf y) = \frac{1}{(2\pi)^{d/2}|\Sigma|}\exp\left( -\frac{1}{2} (\mathbf{y}-\mu)^T \Sigma^{-1} (\mathbf{y}-\mu) \right) := \mathcal N(\mu,\Sigma)
\end{equation}
where $\mathbf y\in \mathbb R^d$, $\mu\in \mathbb{R}^d$ is the mean vector of the random variable (i.e., $\mu_i = \mathbb{E}[Y_i]$), $\Sigma$ is the covariance matrix (i.e., $\Sigma_{i,j}=\mathbb{E}[(Y_i-\mu_i)(Y_j-\mu_j)]$) and $|\Sigma|$ denotes the determinant of $\Sigma$. We remark that the diagonal terms of $\Sigma$ carry the variance of the each component, i.e., $\Sigma_{i,i}=\mathbb{E}[(Y_i-\mu_i)^2]$.

\textbf{Conditioning:} To understand \glspl{gp}, it is crucial for the reader to be acquainted with the concept of conditioning for normal variables. Let us begin by splitting the components of $Y$ into two disjoint sets $A$ and $B$, and decompose the representation of their mean and covariance accordingly as:
\begin{equation}
p([\mathbf y_A, \mathbf y_B]) = \mathcal N \left( \begin{bmatrix}
	\mathbf \mu_A\\
	\mathbf \mu_B
\end{bmatrix}, \begin{bmatrix}
	\Sigma_A, \Sigma_{AB} \\
	\Sigma_{AB}^T, \Sigma_B
\end{bmatrix} \right).
\end{equation}

Now, let us suppose that we \emph{observe} the realization $\mathbf{y}_B$ of the subset $B$ of the random variable $Y$, and we wish to \emph{infer} the value of the remaining, hidden variables $A$. Intuitively, as the correlation between $A$ and $B$ increases, the uncertainty on $A$ after we observe $B$ decreases. This is formalized by the expression of the \emph{posterior} probability $p(\mathbf y_A | \mathbf y_B)$:
\begin{equation} \label{eq:conditional}
	p(\mathbf y_A | \mathbf y_B) = \mathcal N\left(\mu_A \underline{+ \Sigma_{AB} \Sigma_B^{-1}(\mathbf y_B-\mu_B)}, \, \Sigma_A \underline{- \Sigma_{AB} \Sigma_{B}^{-1} \Sigma_{AB}^T} \right).
\end{equation}
Without the observation $\mathbf y_B$, we could have simply computed the distribution of $\mathbf y_A $ alone as $p(\mathbf y_A)= \mathcal N\left(\mu_A, \Sigma_A\right)$. Thus, the underlined terms in \eqref{eq:conditional} are correction factors in the mean and covariance of the variables $A$, brought by the knowledge of $\mathbf y_B$.
%As an extreme case, if the inference is performed at an already observed point $y_A$, then the posterior \eqref{eq:conditional} boils down to a Dirac's delta centered at $y_A$.
Moreover, it is important to observe that in order to compute $p(\mathbf y_A | \mathbf y_B)$, the hardest bit is the \emph{inversion} of the covariance matrix of the observations $\Sigma_{B}$, that can be performed in $\mathcal O((\mathrm{size \ of \ }\mathbf y_B)^3)$.

\textbf{Noisy observations:} Suppose now that the observed samples $\mathbf y_B$ are \emph{noisy}, i.e., they equal the true value of the realization plus an additive Gaussian noise with variance $\sigma^2$. Moreover, let us assume that the noise is \emph{independent} across different components of the variables in $B$. 

Since noise is itself a Gaussian variable then the resulting (noisy) random variable is still Gaussian. Moreover, by independence, the covariance $\Sigma$ stays unchanged, except for its diagonal terms that increase by a  $\sigma^2$. Therefore, the case of noisy samples is accounted by maintaining the same formalism as above and just adding the noise term to the diagonal term of the covariance matrix, i.e., $\Sigma \leftarrow \Sigma + I\sigma^2$.

\subsection{\glspl{gp} for \gls{rrm} problems} \label{sec:GP}
Formally speaking, a \gls{gp} is an uncountable collection of random variables, such that any \emph{finite} collection of those random variables has a multivariate Gaussian distribution. In our case, we will assume that the \emph{unknown performance function $f(.)$ that the controller wants to maximize is modeled as a Gaussian process}. Such an abstract concept leads in our practical case to the following situation. Assume that in the past, the controller has deployed in the system the parameter configurations $\xb(1),\dots,\xb(n)$ and it has observed the corresponding performance metrics:
\begin{equation}
\mathbf{o}(n)= \left[\widetilde{f}(\xb(1)),\dots,\widetilde{f}(\xb(n))\right]^T
\end{equation}
where we recall that $\widetilde{f}(.)$ is a noisy version of the actual (mean) performance metric $f(.)$. Then, by assumption, one can \emph{infer} the performance function at any other value $\xb$ via the \gls{gp} \emph{posterior} probability:
\begin{align}
& p\left(f(\xb)|\mathbf{o}(n)\right) = \nonumber\\ &\qquad 
\mathcal N\left(\mu_f + \Sigma_{f\!\mathbf{o}} \Sigma_{\mathbf{o}}^{-1}(\mathbf{o}(n)-\mu_{\mathbf{o}}), \, \Sigma_f - \Sigma_{f\!\mathbf{o}} \Sigma_{\mathbf{o}}^{-1} \Sigma_{f\!\mathbf{o}}^T \right) \label{eq:posteriorGP}
\end{align}
where the equation above mimics the posterior probability for a Gaussian variable in \eqref{eq:conditional}. 

One can already appreciate that \glspl{gp} have the power to infer the value of the performance function for configurations that have \emph{never} been deployed in the system. This feature really distinguishes \gls{bogp} from other approaches and it will reveal itself to be fundamental to avoid poor \gls{rrm} decisions and converge ``safely'' to an optimal configuration.

Yet, in order to use effectively the posterior \eqref{eq:posteriorGP}, we still have to define the mean vector $\mu$ and covariance matrix $\Sigma$ that we compute via the so-called prior mean and covariance functions that we describe next. 

\textbf{Prior mean function:} To define the mean vectors $\mu_f,\mu_\mathbf{o}$ we resort to the concept of \emph{prior mean function} $m(.):\mathcal X\rightarrow \mathbb R$, that defines for each \gls{rrm}{} configuration $\xb$ the controller's prior \emph{belief} $m(\xb)$ of the performance of the system. 

Ideally, one would like $m(.)\approx f(.)$. By default, if no prior belief is available, one can safely set $m=0$ everywhere. In practice, such belief can be derived by crunching historical data, performing simulations, and/or interrogating human experts with domain knowledge. Another possibility is to define $m(.)$ via a parametric function \cite{williams2006gaussian}, similarly to the covariance function as described next. This would however introduce a further complication that we do not deem necessary for our purposes. Hence, we will just assume $m(.)$ as pre-determined and fixed. 
We provide further intuitions in Section~\ref{sec:ULPC}, where we apply this framework to a practical \gls{rrm} use case.

\begin{figure}
	\centering
	\includegraphics[width=\linewidth]{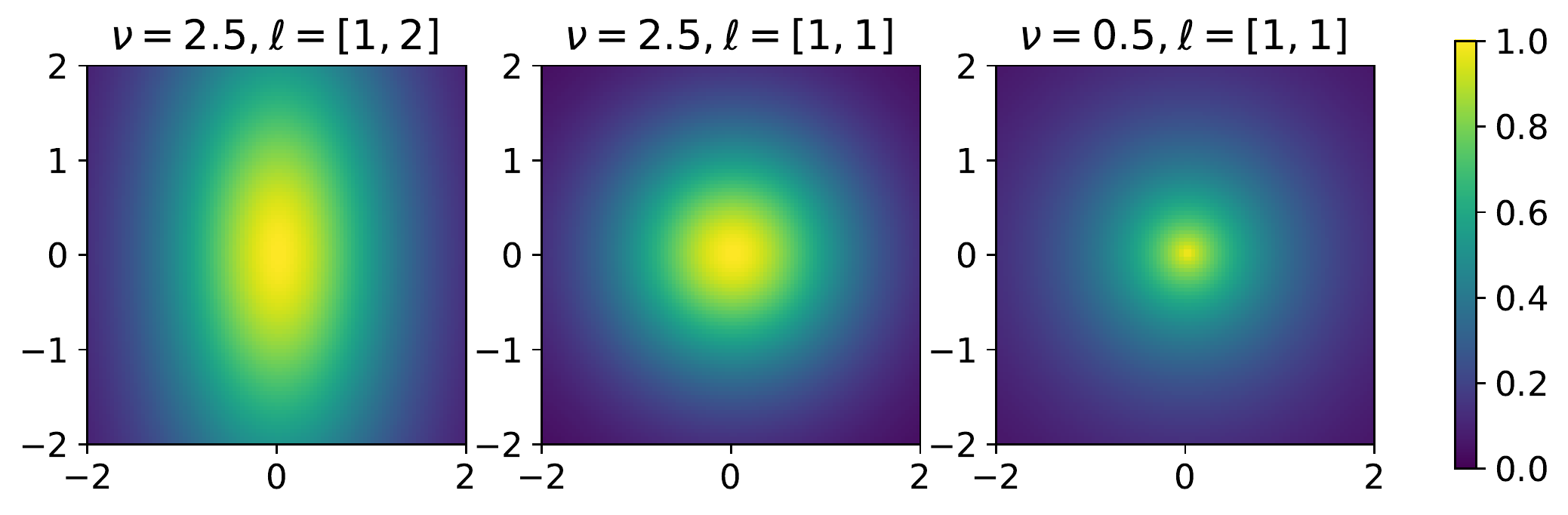}
	\caption{\small \textbf{Mat\'ern kernel.} Illustration of $K^M_{\theta}(\xb,0)$ for different values of the hyperparameters $\theta=[\nu,\ell]$, where $K^M$ is the Mat\'ern kernel \eqref{eq:matern} and $\xb\in\mathbb R^2$. The parameter $\nu$ controls the smoothness of the learned function, and via $\ell$ one can scale differently the control variables.}
	\label{fig:matern}
\end{figure}

\textbf{Covariance function:} To define the covariance matrices $\Sigma_f,\Sigma_{\mathbf{o}},\Sigma_{f\!\mathbf{o}}$ in the \gls{gp} posterior \eqref{eq:posteriorGP}, we construct a function $C(.,.):\mathcal X\times\mathcal X\rightarrow \mathbb R$ that, informally speaking, describes how close we expect system performances obtained by deploying parameters $\xb$ and $\xb'$ to be. Ideally, one would like $C(\xb,\xb')\approx \mathrm{Cov}\big(f(\xb),f(\xb')\big)$.
In practice, $C(.,.)$ is usually defined via the so-called \emph{kernel} functions $K_{\theta}$ that depend on the hyperparameters $\theta$. 
A basic example of kernel function is the Radial Basis Function (RBF), decreasing exponentially with respect to the distance between two candidate points, i.e.:
\begin{equation}
K^{RBF}_\theta(\xb,\xb') := a \exp\left(-\frac{\| \xb-\xb' \|}{2 b^2}\right).
\end{equation}
The set of kernel hyperparameters is $\theta=[a,b]>0$, where $a,b$ regulate the amplitude and the smoothness of the objective function, respectively. The Mat\'ern kernel is more sophisticated and writes:
\begin{equation} \label{eq:matern}
K^{M}_\theta(\xb,\xb') := \frac{1}{\Gamma(\nu)2^{\nu-1}} \left( \sqrt{2\nu} \, d_\ell(\xb,\xb')\right)^\nu \kappa_\nu \left(\sqrt{2\nu} \, d_\ell(\xb,\xb') \right)
\end{equation}
where $d_\ell(\xb,\xb')=\sqrt{\sum_i (x_i-x'_i)^2/\ell_i^2}$, $\theta=[\nu,\ell]$, $\kappa_\nu$ is the modified Bessel function and $\Gamma$ is the Gamma function. Importantly, the parameter $\nu$ controls the smoothness of the learned function and one can scale differently the control variables via $\ell$, as shown in Fig.~\ref{fig:matern}. For a more comprehensive list of kernel functions we refer to \cite{williams2006gaussian}. 
Then, by assuming that the noise affecting the observations is independent across different samples, the covariance function $C(.,.)$ is computed by adding to $K(.,.)$ the observation noise variance term $\sigma^2$ only if $\xb=\xb'$, i.e.,
\begin{equation} \label{eq:cov}
C_{\theta,\sigma}(\xb,\xb') = K_\theta(\xb,\xb') + \sigma^2 \ind(\xb=\xb').
\end{equation}

In practical \gls{rrm} applications, one should choose: i) the kernel function that is believed to suit best the shape of the system performance, and a reasonable initial value for both ii) the kernel hyperparameters $\theta$ and iii) the observation noise variance $\sigma^2$. Concerning i), we believe that the Matérn kernel \eqref{eq:matern} is sufficiently flexible for a large range of functions. Items ii) and iii) are greatly facilitated by the possibility of tuning the hyperparameters in real-time to make the covariance function ``adhere'' at best to the observations in a maximum-likelihood sense, as described in Section~\ref{sec:hyparam_tuning}.

\textbf{Computing the GP posterior \eqref{eq:posteriorGP}:} Once the prior mean $m$ and covariance function $C$ are defined, one can plug them in the expression of the \gls{gp} posterior \eqref{eq:posteriorGP} as follows. We first define  $\mu_f=[m(\xb)]$ as the prior mean for the point $\xb$ at which performance is inferred. The prior mean vector at observed points $\mathbf{o}(n)$ is $\mu_{\mathbf{o}}=[m(\xb(1)),\dots,m(\xb(1))]^T$. The covariance matrix $\Sigma_{\mathbf{o}}$ of the tested \gls{rrm} configurations is such that $[\Sigma_{\mathbf{o}}]_{i,j}=C_{\theta,\sigma}(\xb(i),\xb(j))$ for $i,j=1,\dots,n$. The (mono-dimensional) variance of the system performance $f(\xb)$ to be inferred at the yet-to-be-tested \gls{rrm}{} configuration $\xb$ writes $\Sigma_f=C_{\theta,\sigma}(\xb,\xb)$. Finally, $\Sigma_{f\!\mathbf{o}}$ denotes the covariance column vector between $f(\xb)$ and the observations $\mathbf{o}(n)$, namely $[\Sigma_{f\!\mathbf{o}}]_i=C_{\theta,\sigma}(\xb,\xb(i))$.

\begin{figure*}
	\centering
	\includegraphics[width=\linewidth]{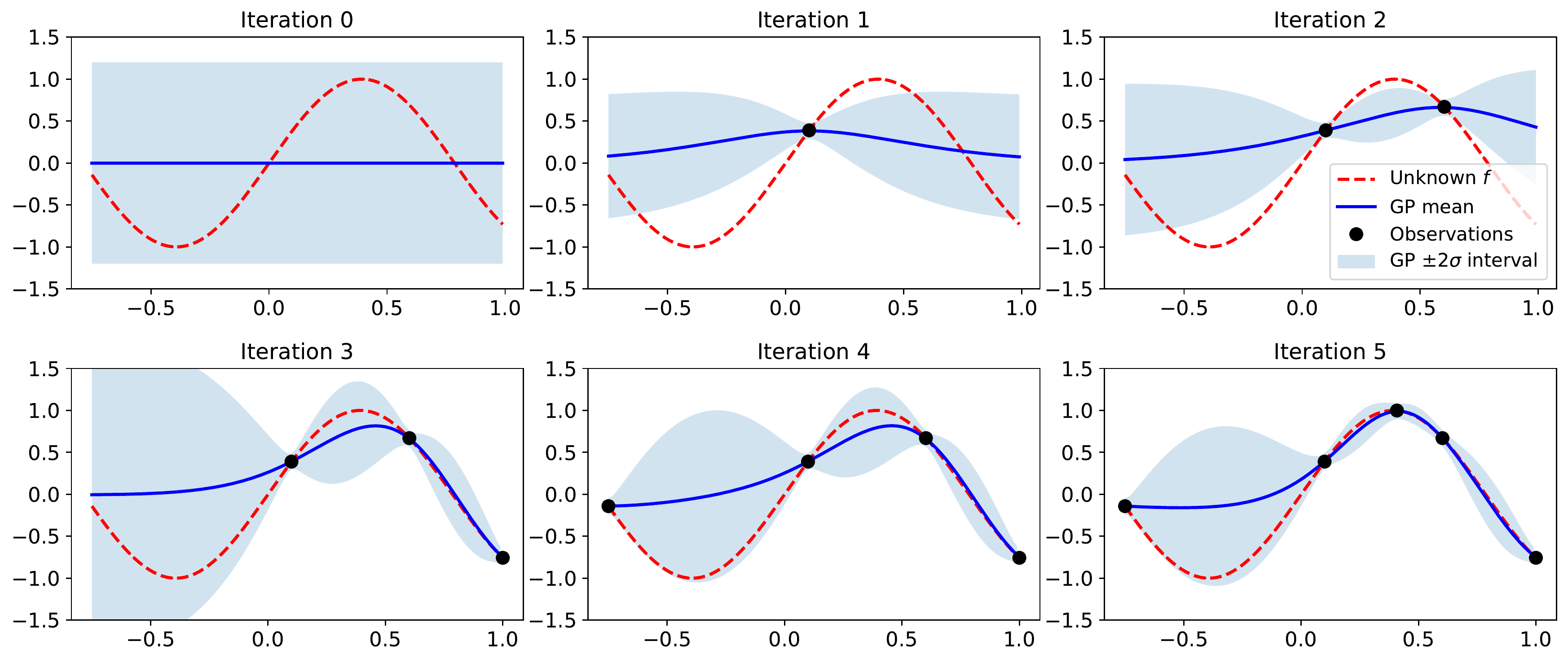}
	\caption{\small \textbf{Example of \gls{bogp}.} The horizontal axis reports the value of the one-dimensional variable $\xb\in \mathbb{R}$ to be optimized. The light blue shaded region and the blue line denote the confidence interval and the mean associated to the GP posterior computed as in \eqref{eq:posteriorGP}, respectively. The red dashed line describes the true function $f(.)$. Black dots are the (noiseless) observations. A Mat\'ern kernel $K^M_{\theta}$ is used (see Eq.~\eqref{eq:matern} and also Fig.~\ref{fig:matern}). The GP starts at iteration 0 with mean prior $m=0$ everywhere.}
	\label{fig:gpex}
\end{figure*}

\subsection{\glspl{gp} for dynamic \gls{rrm} problems} \label{sec:timevarying} 

The bulk of the literature on \gls{bogp} deals with the optimization of an unknown and noisy, yet \emph{static}, function. This scenario fits well with our general \gls{rrm} problem formulation in Section~\ref{sec:bogp}. However, in many practical situations, the system performance function to be optimized changes over time $t$, due to hidden environmental changes in the medium-long term, such as user spatial distribution and traffic intensity. To model this situation, in our dynamic \gls{rrm} problem formulation in Section \ref{sec:bogp} we employed the concept of the hidden network ``state'' $\theta(t)$ that is assumed to evolve exogenously and to be independent of the controller's \gls{rrm} strategy. Then, the performance function $f(.,.)$ itself is assumed to be dependent on both the \gls{rrm} parameters $\xb$ \emph{and} the hidden state $\theta(t)$. In order to be consistent with the \gls{gp} model, we discretize time and we call $t_n$ the time instant at the $n$-th time step. 

\textbf{Augmented observations:} In order to model and track the time-varying function $f(\xb,\theta(t))$ via the \gls{gp} formalism, we follow an approach similar to the one proposed recently in \cite{nyikosa2018bayesian}. We redefine our GP as a function of an augmented variable $\db$ also including the time stamp of each observation:
\begin{equation}
	\db(n):=[\xb(n),t_n], \qquad \forall\, n\ge 1.
\end{equation} 
As we show next, this procedure lets the correlation between two samples depend also on the their time lag, hence allowing one to leverage seasonal patterns that may affect the hidden state $\theta$ to efficiently learn from past samples.

\textbf{Covariance function:} We now define a new GP whose covariance is decoupled as the product between the usual covariance function $C$, only depending on the ``distance'' between two RRM configuration, and a kernel function $K_{\theta^1}$ that only depends on the time lag between two samples. Then, the new covariance function writes:
\begin{align}
    C_{\theta^\delta,\sigma}(\db,\db') = C_{\theta,\sigma}(\xb,\xb') \, \cdot& \, K_{\theta^1}(t,t'), \notag \\
    & \forall \, \xb,\xb'\in \mathcal X, \ \forall\, t,t'\ge 0
\end{align}
where $\db=[\xb,t]$, $\db'=[\xb',t']$ and $\theta^\delta=[\theta,\theta^1]$. An interesting aspect of such a formulation for the \gls{rrm} scenario is that the state $\theta$, although hidden and not always easily identifiable, often follows a \emph{periodic} pattern which is usually strongly correlated with, e.g., the hour of the day or the day within the week. This holds especially if $\theta$ is influenced by the user activity and its geographic position. In such a case, a good choice for $K_{\theta^1}$ is the so-called Exp-Sine-Squared (aka \emph{periodic}) kernel:
\begin{equation} \label{eq:ESS}
	K^{ESS}_{\theta^1}(t,t') = \exp \left( - \frac{2\sin^2\left(\pi |t-t'|/D \right)}{\ell_{ESS}^2} \right)
\end{equation}
where $\theta^1=[D,\ell_{ESS}]$ is the vector of hyperparameters, $D$ describes the periodicity of the hidden state $\theta$ and the \emph{length-scale} $\ell_{ESS}$ regulates the covariance between samples whose time lag is \emph{not} a multiple of the period $D$.

%when at time $n$ the performance $f(\xb(n))$ is observed, the \gls{gp} is fed with all the new $n$ (performance, recency) tuples of the kind $\{\left( f(\xb(n-i)),i \right)\}_{i=0,\dots,n-1}$. Note that this is in contrast to the classic static \gls{gp} framework, where at time $n$, only one new observation $f(\xb(n))$ is added to the record $\mathbf{o}$. Then, at step $n$, the controller's augmented observation dataset comprises all the $n(n+1)/2$ tuples of the kind:
%\begin{equation} \label{eq:obs_dynamic}
%\ob^{\delta}(n):=\Big[\Big(\widetilde f(\xb(i)),\delta\Big) \Big]^T_{i=1,\dots,n, \ \delta=0,\dots,i-1}
%\end{equation}
%where the superscript $\delta$ indicates the dependency on the recency.
%Once the observations $\mathbf{o}^\delta_n$ are collected, similarly to \eqref{eq:posteriorGP}, we aim to infer the system performance at time step $n$, that writes:
%\begin{align}
%	p\left(f(\xb)|\mathbf{o}^{\delta}(n)\right) = \mathcal N\Big(\mu^{\delta}_f & + \Sigma^{\delta}_{f\!\mathbf{o}} {\Sigma^{\delta}_{\mathbf{o}}}^{-1}\left(\mathbf{o}^\delta(n)-\mu^{\delta}_{\mathbf{o}}\right), \notag\\& \, 
%	\Sigma^{\delta}_f - \Sigma^{\delta}_{f\!\mathbf{o}} {\Sigma_{\mathbf{o}}^{\delta}}^{-1} {\Sigma^{\delta}_{f\!\mathbf{o}}}^{T} \Big). \label{eq:posteriorGP_dynamic}
%\end{align}
%As in the static case, the mean and covariance matrices $\mu^{\delta},\Sigma^{\delta}$ are computed with the aid of the prior mean and covariance functions $m$ and $C$, respectively, by augmenting them with the temporal dependency as described next.

\textbf{Mean prior function:} Since the objective function $f$ is supposed to vary over time, it is reasonable to let the prior mean function $m$ depend on time as well. Then, $m(\xb,t)$ is the system performance that one expects if the \gls{rrm} configuration $\xb$ is deployed at time instant $t$. Yet, as in the static case, such mean prior should be considered as a bonus: if such information is not available, one can settle for a \emph{constant} prior mean $m(\xb,t)=c$, for some arbitrary constant $c$.

\textbf{Computing the \gls{gp} posterior:} Once the observations $\mathbf{o}(n)$ are collected, we can infer the performance of any \gls{rrm}{} configuration $\xb$ at the next deployment time $t_{n+1}$ as:
\begin{align}
	p\Big(f(\xb,\theta(t_{n+1}))|\mathbf{o}(n)\Big) = \mathcal N\Big(\mu^{\delta}_f & + \Sigma^{\delta}_{f\!\mathbf{o}} {\Sigma^{\delta}_{\mathbf{o}}}^{-1}\left(\mathbf{o}(n)-\mu^{\delta}_{\mathbf{o}}\right), \notag\\& \, 
	\Sigma^{\delta}_f - \Sigma^{\delta}_{f\!\mathbf{o}} {\Sigma_{\mathbf{o}}^{\delta}}^{-1} {\Sigma^{\delta}_{f\!\mathbf{o}}}^{T} \Big) \label{eq:posteriorGP_dynamic}
\end{align}
where $\mu_f^\delta=[m(\xb,t_{n+1})]$ is the prior mean at point $\xb$ and at time $t_{n+1}$ and the vector of the observation means is $\mu_{\mathbf o}^\delta=\left[m\big(\xb(1),t_1\big),\dots,m\big(\xb(n),t_n\big)\right]^T$. 
% We are finally ready to define the mean and covariance matrices appearing in the \gls{gp} posterior expression \eqref{eq:posteriorGP_dynamic} for dynamic functions. For ease of notation, we denote by $\xb(\mathbf{o}_i^\delta)$ and $\delta(\mathbf{o}_i^\delta)$ the \gls{rrm}{} configuration and the recency associated with the $i$-th observation $\mathbf{o}^\delta_i(n)$, respectively, according to the ordering defined in \eqref{eq:obs_dynamic}. 
%We then denote $\mu_f^\delta=[m(\xb,n)]$ as the prior mean at point $\xb$ and at time $n$. The vector of the observation means is $\mu_{\mathbf o}^\delta=\left[m\big(\xb(\mathbf{o}_i^\delta),n-\delta(\mathbf{o}_i^\delta)\big)\right]^T_{i=1,\dots, n(n+1)/2}$. 
The covariance matrix $\Sigma^\delta_{\mathbf{o}}$ of the tested \gls{rrm}{} configurations is such that $[\Sigma^\delta_{\mathbf{o}}]_{i,j}= C_{\theta^\delta,\sigma}(\db(i),\db(j))$, whereas the (mono-dimensional) variance of the system performance $f(\xb)$ to be inferred at the yet-to-be-tested \gls{rrm}{} configuration $\xb$ is $\Sigma^\delta_f=[C_{\theta^\delta,\sigma}([\xb,t_{n+1}],[\xb,t_{n+1}])]$. 
Finally, $\Sigma^\delta_{f\!\mathbf{o}}$ is the covariance column vector between $f(\xb,\theta(t_{n+1}))$ and the observations, namely $[\Sigma_{f\!\mathbf{o}}]_i=C_{\theta^\delta,\sigma}\big([\xb,t_{n+1}],\db(i)\big)$.

\subsection{Hyperparameter tuning} \label{sec:hyparam_tuning}

We mentioned above that the covariance function $C$ is parametrized by the kernel hyperparameters $\theta$ ($\theta^\delta$, in the dynamic case; yet in this section we will stick to the static \gls{rrm}{} notation for simplicity) and the observation noise variance $\sigma$. These are generally \emph{unknown} to the controller, which has at best some prior belief $\theta_0$ and $\sigma_0$ on them. The \gls{gp} framework offers a principled way to tune $\theta,\sigma$ in real-time, as new system performance observations $\mathbf{o}(n)$ are gathered. This is usually performed by finding those $\theta,\sigma$ that maximize the likelihood of the observed function values $\mathbf{o}(n)$, i.e.,
\begin{equation} \label{eq:opt_hyperparameters}
	\argmax_{\theta,\sigma} \, p\big( \widetilde{f}(\xb(1)),\dots,\widetilde{f}(\xb(n)) \big) = \mathcal N(\mu_{\mathbf{o}},\Sigma_{\mathbf{o}})
\end{equation}
where $\mu_{\mathbf{o}}$ and $\Sigma_{\mathbf{o}}$ are the mean prior and covariance of the observations $\mathbf{o}(n)=[\widetilde f(\xb(1)), \dots,$ $\widetilde f(\xb(n))]^T$, respectively.

We remark that \eqref{eq:opt_hyperparameters} is generally a non-convex problem, and one usually looks for a local maximum. Bayesian model-selection techniques for this goal are presented in \cite{williams2006gaussian}, Chapter 5. More recent advances are proposed in  \cite{blum2013optimization}, \cite{pautrat2018bayesian}. Otherwise, more standard gradient-descent optimization techniques can be used for \eqref{eq:opt_hyperparameters}, such as the Broyden-Fletcher-Goldfarb-Shanno (BFGS) algorithm \cite{avriel2003nonlinear}, which is suited to general unconstrained and non-convex optimization problems.

\subsection{Dealing with a large number of observations}

The main computational bottleneck of the \gls{gp} inference engine resides in the inversion of the covariance matrix of the observations, called  $\Sigma_{\mathbf o}$ and $\Sigma^\delta_{\mathbf o}$ in the static and dynamic scenarios respectively, and used in \eqref{eq:posteriorGP},  \eqref{eq:posteriorGP_dynamic} for the computation of the \gls{gp} posterior. 
At time step $n$, the matrix inversion requires $\mathcal O(n^3)$ computations.

There exist methods in the literature to alleviate such issues, see, e.g., \cite[Ch. 8]{williams2006gaussian}. These include two large classes of methods: i) approximating the covariance matrix as a matrix with reduced rank $n'$, which allows one to use the Woodbury matrix identity, requiring the inversion of an $n'$-by-$n'$ matrix and ii) reducing the size of the set of observations by picking those maximizing a pre-defined significance criterion. Moreover, in the case of dynamic \gls{rrm}{}, a sensible choice is to include in the observation set $\ob(n)$ only the outcomes of the last $n'<n$ tested \gls{rrm} configurations, where $n'$ should depend on the expected variation speed of the hidden state.

It is also worth mentioning that in the case of static \gls{rrm} problems, the controller often cannot afford to explore among a large number of different configurations before sticking to a definitive choice. Thus, if the maximum envisioned number of tested configurations is less than $10^3$, the matrix inversion can be easily performed, e.g., via Cholesky decomposition \cite{williams2006gaussian}.

\section{\gls{bo} for \gls{rrm} problems} \label{sec:BOdetails}

The \gls{gp} framework offers a powerful inference engine allowing one to predict, given the historical observations $\ob(n)$, the current value of the system performance function for any possible \gls{rrm} configuration. This is achieved via the posterior distribution formula \eqref{eq:posteriorGP}.\footnote{In this section we will stick to the static \gls{rrm} formulation. However, everything translates easily to the dynamic setting by just considering \eqref{eq:posteriorGP_dynamic} as the \gls{gp} posterior and by fixing the inference time to $t_{n+1}$.}

At each step $n+1$, the role of \gls{bo} is to leverage the \gls{gp} inference based on observations $\ob(n)$ and choose the next \gls{rrm} configuration $\xb(n+1)$ to be deployed in the system.
We first observe that the \gls{bo}'s task is of strategic nature, and it is intimately connected with the Markov Decision Process (MDP) \cite{puterman2014markov} control framework. 
In fact, if we define the control ``state'' as the information necessary to take the next decision $\xb(n+1)$---in our case, the \gls{gp} posterior \eqref{eq:posteriorGP} itself---then the controller alternates between i) taking a decision $\xb(n+1)$ based on the current ``state'', ii) observing the associated random ``reward'' $\widetilde{f}(\xb(n+1))$ and iii) experiencing a deterministic transition to a new ``state'' via the \gls{gp} posterior formula. 
This procedure matches the MDP state-reward-transition formulation; however, the classic MDP algorithms (e.g., policy iteration, value iteration \cite{puterman2014markov}, or even more recent neural approximations \cite{powell2007approximate}) are highly impractical in this scenario due to the astronomical size of the associated state space. 

For such reason, \gls{bo} takes a computationally simpler approach and optimizes at each step $n+1$ a so-called \emph{acquisition function} $u(,|\mathbf{o}(n))$ that still depends on the previous observations via the \gls{gp} posterior \eqref{eq:posteriorGP}. 
The next selected \gls{rrm}{} configuration $\xb(n+1)$ is the one maximizing the acquisition function $u$:
\begin{equation} \label{eq:BO}
	\xb(n+1) = \argmax_{\xb\in \mathcal X} u\big(\xb|\ob(n)\big).
\end{equation}
The function $u$ should account for the exploitation-vs.-exploration dilemma that naturally arises in such situations. In fact, the controller would want to sample a point with high \emph{expected} performance, while minimizing the \emph{risk} that the \emph{actual} performance drops too low.

Although in the literature there exist several ways to define the acquisition function $u$ (e.g., see \cite{theodoridis2015machine}), we will illustrate three relevant approaches for our purposes, called \emph{expected improvement} and \emph{knowledge gradient} and \emph{upper confidence bound}. The first two ensure that the next configuration $\xb(n+1)$ in \eqref{eq:BO} is optimal for the MDP described above over a truncated time horizon, but under different assumptions on the observation noise.

\textbf{Expected Improvement (EI):} The EI is arguably the most widely known acquisition function. It was first introduced by \cite{movckus1975bayesian} in a general Bayesian setting, and then applied to \glspl{gp} in \cite{jones1998efficient,huang2006sequential}. 
To understand the theoretical motivations behind the EI, it is useful to perform the following thought experiment. Suppose that i) the observations are noiseless, i.e., $\widetilde f(.)=f(.)$, ii) we are at time $n$, iii) at time $n+1$, we can test one more \gls{rrm} configuration $\xb(n+1)$, iv) at time $n+2$, we have to stick to one of the configurations already tested $\xb(1),\dots,\xb(n+1)$, and v) the experiment ends. Thus, knowing that at time $n+2$ we can always select the best point so far, the performance at time $n+2$ is $\max\left\{f(\xb(n+1)), \ \max_{i=1,\dots,n}  f(\xb(i))\right\}$. Therefore, the \gls{rrm}{} choice $\xb(n+1)$ that maximizes the expected performance at time $n+2$ is the one maximizing the expected improvement $u^{\mathrm{EI}}$ with respect to the best configuration so far, where:
\begin{align}
	u^{\mathrm{EI}}(\xb|\ob(n)) = & \, \mathbb{E} \left[ f(\xb) - \max_{i=1,\dots,n}  f(\xb(i)) \, \Big| \, \mathbf{o}(n) \right]^+. \label{eq:EI}
\end{align}
Note that $[.]^+=\max(.,0)$ is the positive part and the expected value is with respect to the \gls{gp} posterior \eqref{eq:posteriorGP}.
A reason for its popularity is that $u^{\mathrm{EI}}$ has a closed form for GPs that writes:
\begin{align}
u^{\mathrm{EI}}(\xb|\ob(n)) = & \, \left( \mathbb E\left[f(\xb)|\ob(n)\right]-\max_{i=1,\dots,n} f(\xb(i)) -\xi\right) \Phi(Z) + \notag \\ & + \mathrm{Std} \left[f(\xb)|\ob(n)\right] \varphi(Z), \qquad \forall \, \xb\in \mathcal X 
\end{align}
where $\mathbb E\left[f(\xb)|\ob(n)\right]$ and $\mathrm{Std} \left[f(\xb)|\ob(n)\right]$ are the mean and standard deviations of the \gls{gp} posterior, respectively; $\Phi$ and $\varphi$ are the cumulative and probability density function of the standard normal distribution, respectively, and $Z$ is defined as:
\begin{equation}
Z = \frac{\mathbb E\left[f(\xb)|\ob(n)\right] - \max_{i=1,\dots,n} f(\xb(i)) - \xi}{\mathrm{Std} \left[f(\xb)|\ob(n)\right]}.
\end{equation}
Importantly, $\xi$ denotes the amount of exploration during optimization. A risk-sensitive \gls{rrm} controller will choose $\xi$ small (e.g., $\xi=0.01$ or even $\xi=0$) to minimize the risk of a performance drop during optimization. In fact, the EI offers a natural exploitation-vs.-exploration interpretation, since one wants to sample a point with a good trade-off between high expected performance (cf. exploitation) and low uncertainty (cf. exploration). 

In practice, to solve \eqref{eq:BO} for the EI case, one can use first- or second-order methods. E.g., one suitable technique is to calculate first derivatives and use the quasi-Newton method L-BFGS-B as in \cite{liu1989limited}.

It is important to remark that in our general \gls{rrm} scenario, the acquisition function $u^{\mathrm{EI}}$ in \eqref{eq:EI} is \emph{not} well defined. In fact, it depends on the \emph{true} value of the system performance function $f(.)$, while the controller can only actually estimate a noisy version $\widetilde f(.)$ of it. A variety of heuristic approaches exist to deal with noisy observations \cite{frazier2018tutorial}. For instance, one can simply replace $f(.)$ with its noisy observation $\widetilde{f}(.)$, else with the \gls{gp} posterior mean $\mathbb E\left[f(.)|\ob(n)\right]$.

%Moreover, EI is particularly suited when the observations are noiseless, i.e. when $f(.)=\widetilde{f}$, for which it was proven in \cite{movckus1975bayesian} to be the one-step ahead optimal solution of the Bayesian MDP described above, i.e., optimal over a truncated one-step horizon. 
If the observation noise is dominating---especially if the \gls{rrm} configuration is updated at a fast time scale, which does not allow the controller to have sufficient samples to estimate $f(\xb)$ well enough---then the performance of the EI degrades. In this case, one can adopt a different acquisition function that we describe next.

\textbf{Knowledge Gradient (KG):} The KG acquisition was first derived in \cite{frazier2009knowledge} as a variant of EI that is particularly suited to the case of noisy observations of the performance function. KG can be defined in a constructive manner by recycling the thought experiment for EI above while modifying the assumptions i) and iv). We here suppose that i) observations are noisy and iv) at time $n+2$, one can choose \emph{any} \gls{rrm} configuration (even one that has not been tested in the past). 
%Assume that at time $n$ the controller were to stop exploration and stick to a certain \gls{rrm}{} configuration forever, from time $n+1$ onward. Then, it would select the \gls{rrm}{} configuration with the highest posterior mean $\mu^*(n+1)=\max_{x\in \mathcal X} \mathbb E\left[f(\xb)|\ob(n)\right]$. 
Our goal is to choose $\xb(n+1)$ at time $n+1$. If $\xb(n+1)$ is selected with observation $\widetilde{f}(\xb(n+1))$, then at step $n+2$, one would want to achieve the highest expected performance 
\begin{equation}
\mu^*(n+2)=\max_{\xb'\in \mathcal X} \mathbb E\left[f(\xb(n+2)=\xb') \, \big| \, \ob(n),\widetilde{f}(\xb(n+1))\right].
\end{equation}
Equivalently, the controller would seek to maximize the \emph{improvement} of system performance $\mu^*(n+2)-\mu^*(n+1)$, where we analogously define $\mu^*(n+1)=\max_{x\in \mathcal X} \mathbb E\left[f(\xb)|\ob(n)\right]$. Yet, at time $n$, the value of $\widetilde{f}(\xb(n+1))$ is unknown and $\mu^*(n+2)-\mu^*(n+1)$ a random quantity. Hence, we take its expectation to finally compute the KG acquisition function:
\begin{equation} \label{eq:KG}
u^{\mathrm{KG}}(\xb|\ob(n)) = \mathbb{E} \left[ \mu^*(n+2)-\mu^*(n+1) \, \big| \, \xb(n+1)=\xb \right].
\end{equation}

Although KG performs better than EI in the noisy observation scenario, it still suffers from computational issues since there exists no closed-form expression for \eqref{eq:KG} and one has to resort to evaluate it via simulations. We remark that an efficient and scalable approach, based on multi-start stochastic gradient ascent, has been proposed in \cite{wu2016parallel}.

\textbf{Upper Confidence Bound (UCB):} Both EI and KG aim at improving the value of the objective function with respect to the \emph{best} observation so far. Yet, the objective function itself may vary over time with the state $\theta$, as in Section \ref{sec:timevarying}. Thus, at each step $n$ EI and KG should in principle compute the best observation among all those where the network found itself in the same state $\theta=\theta(t_n)$, and define the improvement accordingly. 
This clearly poses a challenge, since the state $\theta$ is generally hidden or hard to infer. For this reason, in the dynamic scenario, it is more convenient to employ the UCB acquisition function \cite{KKS10}, that only depends on the GP posterior and not on previous observations. UCB naturally trades off exploration and exploitation, i.e., choosing points with high GP posterior variance and mean, respectively, via the coefficient $\beta_n\ge 0$:
\begin{equation} \label{eq:UCB} 
u^{\mathrm{}}(\xb|\ob(n)) = \mathbb{E}\left[f(\xb)|\mathbf{o}(n)\right] + \beta_n \, \mathrm{Std} \left[f(\xb)|\mathbf{o}(n)\right].
\end{equation}
The closed form expression of $u^{\mathrm{UCB}}$ directly stems from the GP posterior definitions in \eqref{eq:posteriorGP},\eqref{eq:posteriorGP_dynamic}. We refer to \cite{KKS10} for the tuning of coefficients $\beta_n$'s.

\textbf{Safe exploration and termination:} It should be now apparent that one of the great advantages of the \gls{gp} resides in the possibility of evaluating the risk-vs.-benefit trade-off associated with any possible \gls{rrm} configuration. In turn, via the optimization of the acquisition function, \gls{bo} allows the controller to fully exploit the information provided by the \gls{gp} inference engine to improve the system performance and, remarkably, to avoid to deploy configurations that are inferred to perform poorly with high probability. 

In the static \gls{rrm} scenario, the \gls{bo} framework also allows the controller to naturally figure out when it is reasonable to terminate the exploration and eventually stick to a fixed \gls{rrm}{} configuration. For instance, the exploration may stop when the achievable expected improvement of the system performance is lower than a threshold $\epsilon$, i.e.,
\begin{equation}
\max_{\xb\in\mathcal X} u(\xb|\ob(n))<\epsilon
\end{equation}
where $u:=u^{\mathrm{EI}}$ or $u:=u^{\mathrm{KG}}$. Otherwise, the controller could simply set a maximum number $\overline n$ of exploration steps. Afterwards, the controller would stick to the \gls{rrm} configuration $\xb^*=\argmax_{\xb(i),i=1,\dots,n} \widetilde{f}(\xb(i))$ that allowed the system to achieve the highest performance so far.

We recap the main steps of \gls{bogp} for static and dynamic \gls{rrm} problems in Algorithms \ref{algo:BOGP} and \ref{algo:BOGP_dynamic}, respectively. We show in Fig. \ref{fig:gpex} a visual insight of the BOGP behavior for one optimization variable. In the next section, we illustrate the benefits of adopting the \gls{bogp} framework for the optimization of network-wide system parameters in a specific uplink power control \gls{rrm} use-case.

\begin{algorithm}[h]
	\SetAlgoLined
	\textbf{Initialization.} Set $n=0$\;
	Define a prior mean function $m(.)$. By default, $m=0$\;
	Choose the covariance kernel function $K_\theta(.,.)$\;
	Initialize the hyperparameters $\theta,\sigma$\;
	Define a termination threshold $\epsilon$ and a maximum number of exploration steps $\overline n$\;
	\While{$\max_{\xb\in\mathcal X} u(\xb|\ob(n))\ge\epsilon$ \textnormal{or} $n\le \overline n$}{
		Find the next \gls{rrm}{} configuration $\xb(n+1)$ via \eqref{eq:BO}\;
		Deploy $\xb(n+1)$ in the system and observe the performance $\widetilde f(\xb(n+1))$\;
		Update hyperparameters $\theta,\sigma$ via \eqref{eq:opt_hyperparameters}\;
		Set $n\leftarrow n+1$\;
	}
    Deploy $\xb^*=\argmax_{\xb(i),i=1,\dots,n} \widetilde{f}(\xb(i))$
	\caption{\gls{bogp} for static \gls{rrm} problems}
	\label{algo:BOGP}
\end{algorithm}

\begin{algorithm}[h]
	\SetAlgoLined
	\textbf{Initialization.} Set $n=0$.\;
	Define a prior mean function $m(.)$. By default, $m=0$\;
	Choose the spatial covariance kernel function $K_\theta(.,.)$\;
	Choose the temporal covariance kernel function $K^\delta_{\theta^1}(.)$\;
	Initialize the hyperparameters $\theta^\delta,\sigma$\;
	\While{\texttt{True}}{
		Find the next \gls{rrm}{} configuration $\xb(n+1)$ via \eqref{eq:BO}, with $u:=u^{\mathrm{UCB}}$, see \eqref{eq:UCB}\;
		Deploy $\xb(n+1)$ in the system and observe the performance $\widetilde f(\xb(n+1),\theta(t_{n+1}))$\;
		Update hyperparameters $\theta^\delta,\sigma$ via \eqref{eq:opt_hyperparameters} (by replacing $\theta$ with $\theta^\delta$)\;
		Set $n\leftarrow n+1$\;
	}
	\caption{\gls{bogp} for dynamic \gls{rrm} problems}
	\label{algo:BOGP_dynamic}
\end{algorithm}

\section{Open Loop Power Control} \label{sec:ULPC}
\begin{figure}
\centerline{\includegraphics[width=\columnwidth]{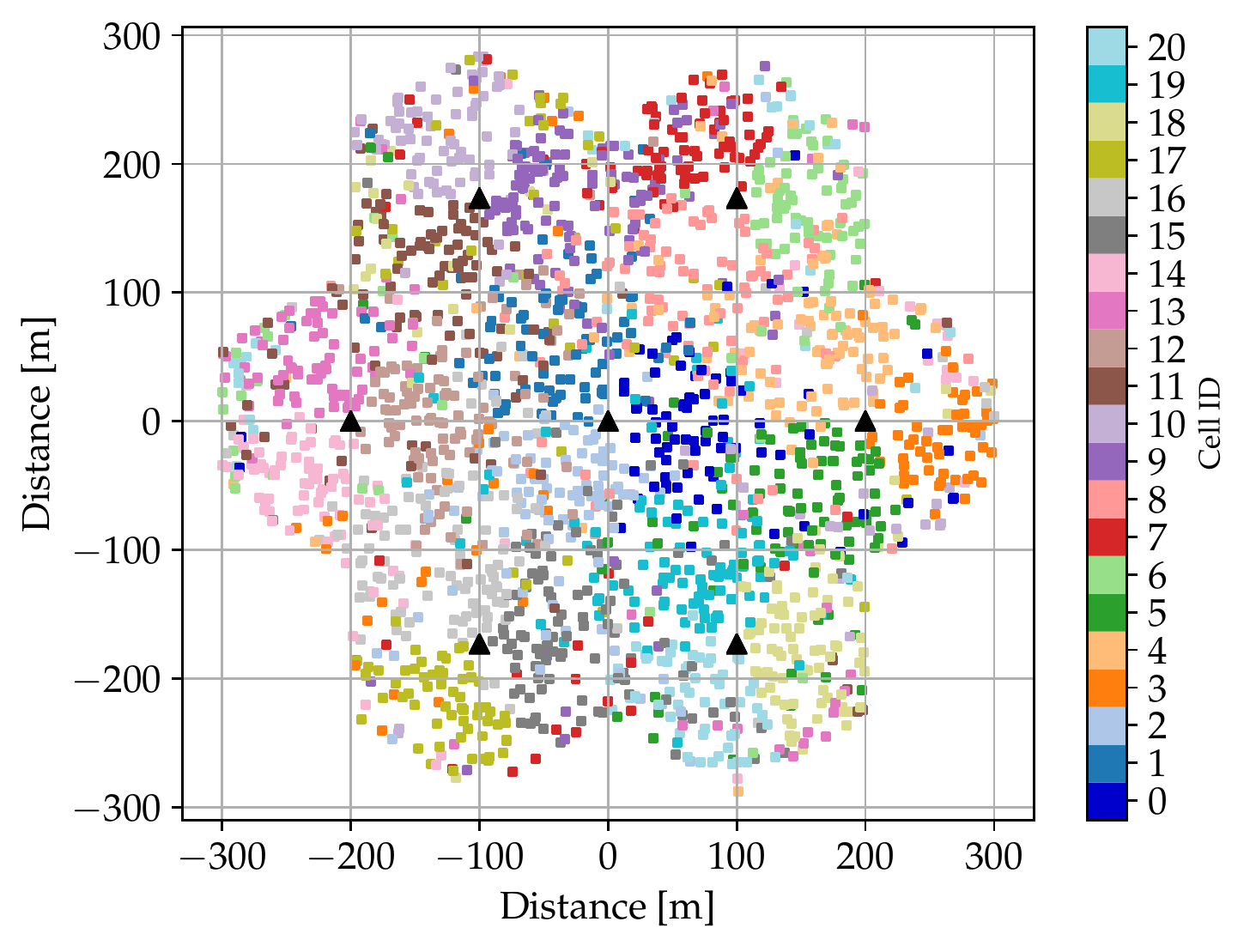}}
\caption{\small \textbf{Network layout.} Triangles denote \gls{bs} sites, squares describe \gls{ue} locations, and their color characterizes their serving cell.}
\label{fig:network}
\end{figure}

\begin{figure}
\centerline{\includegraphics[width=\columnwidth]{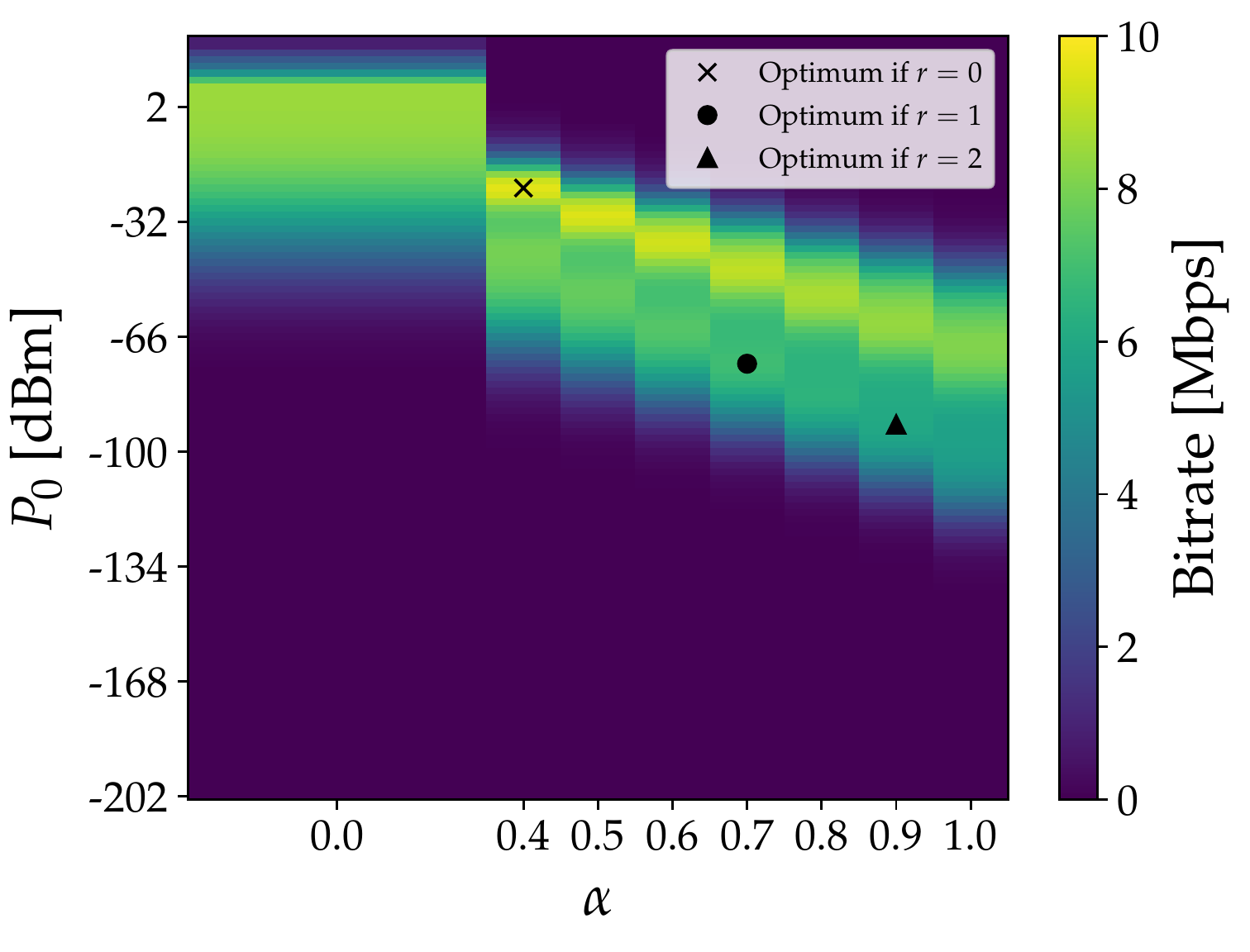}}
\caption{\small \textbf{System performance function.} Utility function $u(\alpha, P_0)$ with fairness $r=0$ on the $(\alpha, P_0)$ solution space when simulated with $K=4$ \glspl{ue}. The optimum $x^*=(\alpha^*, P_0^*)$ for other values of $r$ are also shown.}
\label{fig:target}
\end{figure}

In this section, we consider a practical instance of the general \gls{rrm} problem defined in Section \ref{sec:bogp}, namely \gls{ulpc}, which consists in managing the transmit power of \glspl{ue} in cellular networks.
In \gls{lte} and \gls{5gnr}, the transmissions of different \glspl{ue} within the same cell must reach the \gls{bs} with approximately the same power. This is important to manage the average cell spectral efficiency, as well as user throughput fairness.
For cell-edge \glspl{ue}, the same transmit power spectral density would cause more interference to neighbor cells when compared to cell-center \glspl{ue}.
For this reason, 3GPP adopts the so-called fractional power control approach, where the uplink transmit power compensates only for a fraction of the total pathloss.

As per 3GPP TS 38.213, \glspl{ue} calculate their \gls{pusch} transmission power $P_{PUSCH}$ in dBm as follows:
\begin{align}
P_{PUSCH}=\min\{ P_{CMAX},\  P_0 +10\log(M_{RB}) + \alpha PL + CL \} \label{eq:p_pusch}
\end{align}
where $P_{CMAX}$ is the \gls{ue} configured maximum output power, $M_{RB}$ is the number of \glspl{prb} allocated to the \gls{ue}, $\alpha$ is the fractional power control compensation parameter, $PL$ is the downlink pathloss estimate, $CL$ is the closed-loop power control adjustment and $P_0$ is the power per \gls{prb} that would be received under full pathloss compensation.

\glspl{ue} transmitting with power higher than needed will experience battery drain and cause unnecessary inter-cell interference to the uplink of nearby cells.
On the other hand, \glspl{ue} transmitting with low powers will have a poor Quality of Experience or even worse, connectivity losses.
A proper management of the transmit powers of the uplink data channels is therefore essential for the network to operate well.

Open-loop power control (\Gls{olpc}) refers to the selection of slow-changing parameters that \glspl{ue} can use to autonomously select their own transmit power.
As per \eqref{eq:p_pusch}, the parameters that control this behavior on each network cell are $P_0$ and $\alpha$.
The selection of values (see Table~\ref{tab:olpc_params}) for these parameters defines the uplink inter-cell interference operation point and hence the average uplink radio performance.
\Gls{clpc} then applies fast transmit power corrections around this point.

\begin{table}[htbp]
\caption{Allowed values for the 3GPP \gls{olpc} parameters}
\begin{center}
\begin{tabular}{|r|l|}
\hline
$P_0$& $-202, -200, ..., +22, +24$ \\
\hline
$\alpha$ & $0, 0.4, 0.5, 0.6, 0.7, 0.8, 0.9, 1$\\
\hline
\end{tabular}
\label{tab:olpc_params}
\end{center}
\end{table}

As of 2021, most \glspl{mno} choose a set of default values for the \gls{olpc} parameters and then update individual cells on-demand, following technical intuitions.
However, the inter-cell interference behavior of a cellular network is not intuitive at all, which makes this procedure costly and sub-optimal.

Exhaustive search approaches are simply not scalable due to the combinatorial nature of the problem.
In fact, there are eight possible standard values of $\alpha$ and $114$ possible values for $P_0$, hence, a total of $N=8\cdot114=912$ possible \gls{olpc} configurations for any given cell in a network.
In a network with $C$ cells, there are $N^C$ possible \gls{olpc} network configurations. As an example, this rises to nearly 760 million combinations even for a small network with three cells.
Importantly, some \gls{olpc} configurations yield very low uplink performance (see \cite{Mullner2009}, \cite{Haider2018}).
These are unknown \emph{a priori} and are to be avoided.
Testing these configurations for the sake of exploration is usually unacceptable to \glspl{mno}, which is why methods are needed to optimize the \gls{olpc} configuration in few steps while exploring safely.

As in most \gls{rrm} problems, the relationships between the \gls{olpc} parameters and the uplink \glspl{kpi} are complex and hard to model.
For this reason, we propose to treat this dependence as a black-box and optimize it via a Bayesian approach, with a Gaussian process modeling the performance function based on the preferred system KPI.
A main advantage of this approach is that domain knowledge and/or system simulation results can be exploited to design an initial prior of the performance function, allowing one to significantly increasing the safety and convergence speed of the exploration procedure.
Next we provide details about this approach.
\begin{figure*}[!t]
\centering
\subfloat[\gls{olpc} configurations evaluated by the optimization algorithm.]{\includegraphics[width=.5\linewidth]{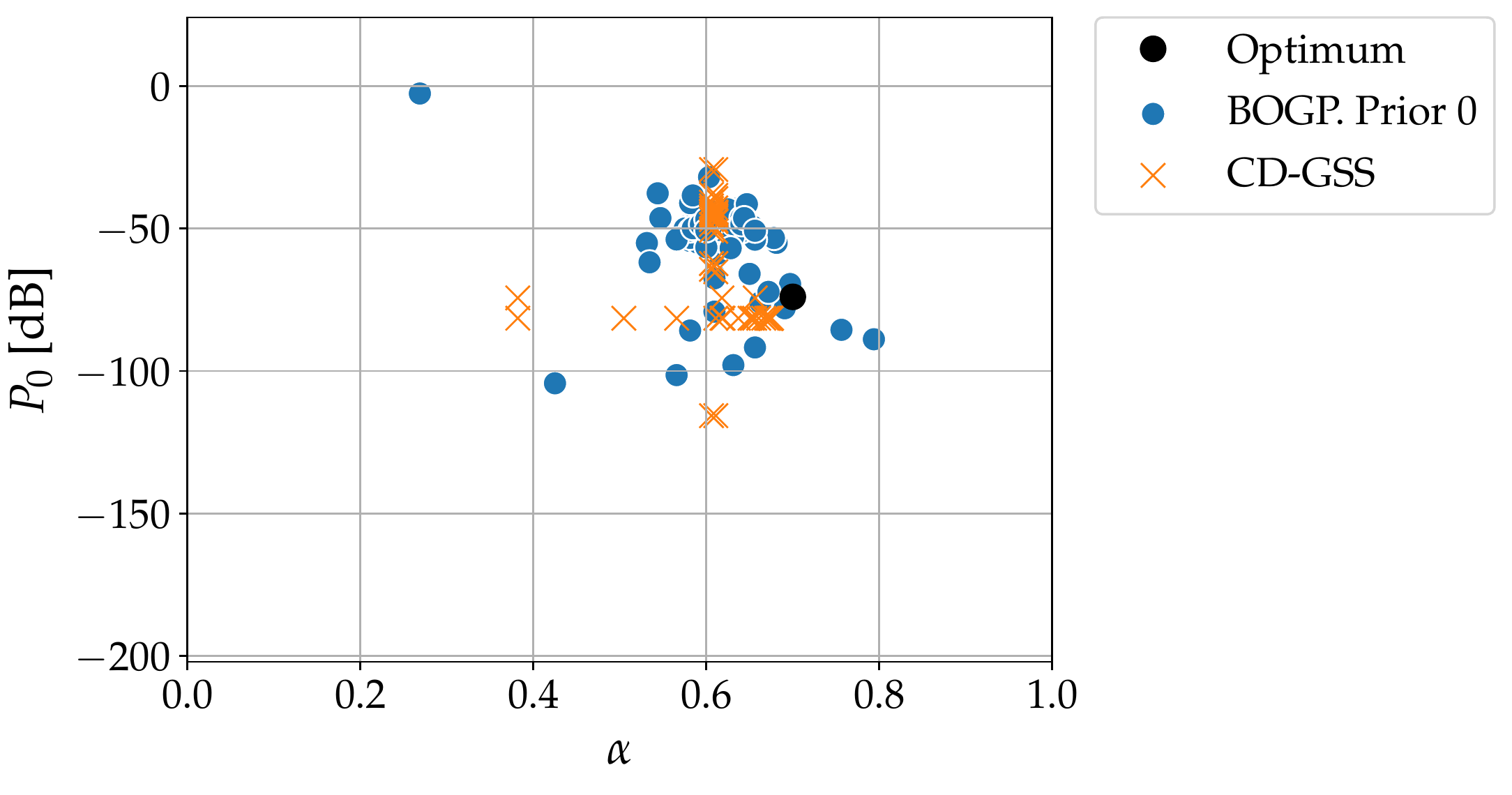}\label{fig:trajectory}}
\hfil
\subfloat[Convergence time comparison. The solid line depicts the mean performance and the shadowed region the 95\% confidence interval.]{\includegraphics[width=.4\linewidth]{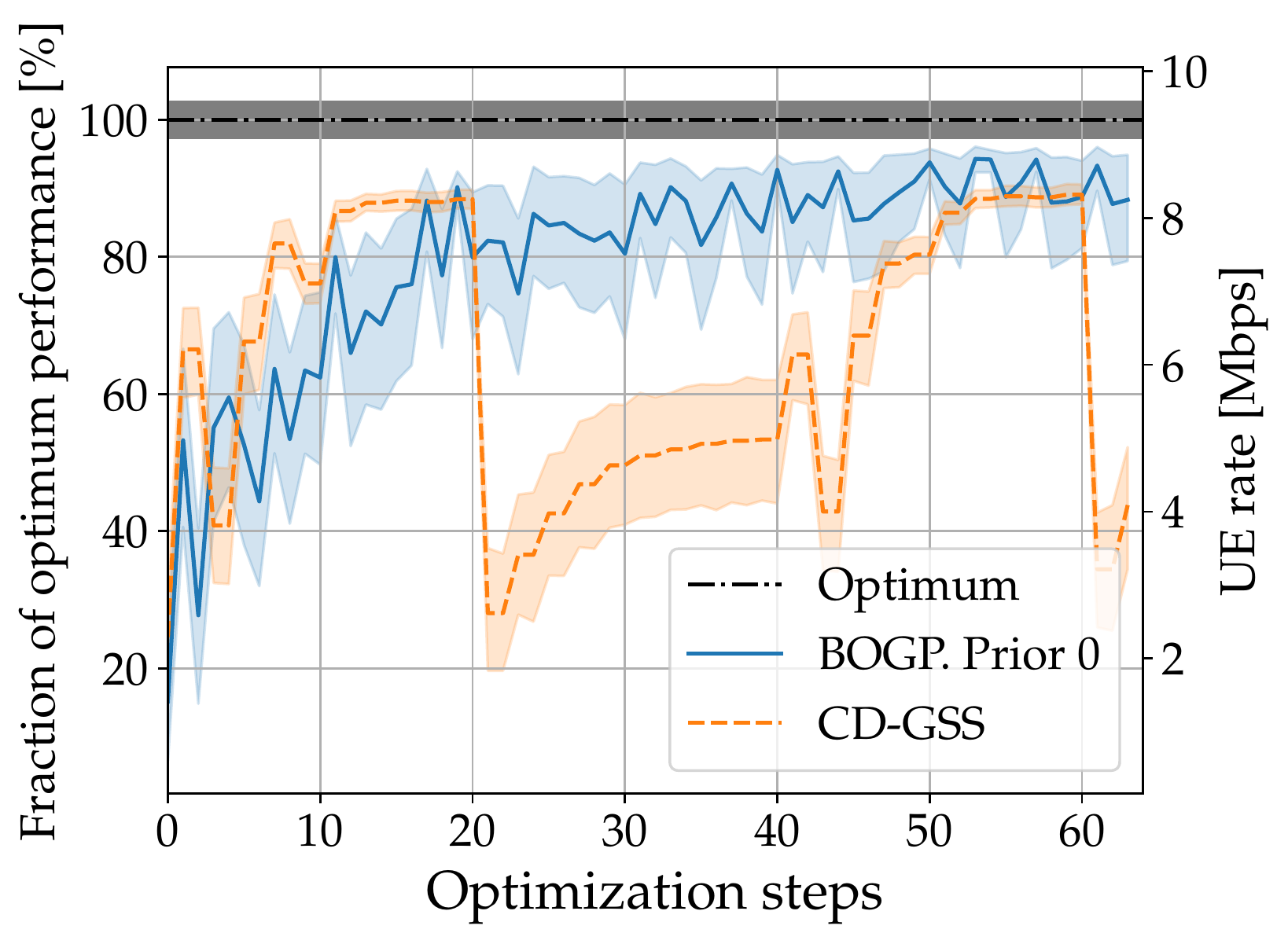}\label{fig:convergence}}
\caption{\textbf{BOGP convergence.} Evaluation with $K=4$ active \glspl{ue} per snapshot and a fairness level of $r=1$ on the network architecture of Fig.~\ref{fig:network}. The $(\alpha, P_0)$ solutions explored by \gls{bogp} with Prior 0 (see \cref{fig:priors_mean}) are shown. The performance of a \gls{cd} algorithm with \gls{gss} on each coordinate is also reported for reference.}
\label{fig:performance}
\end{figure*}

\subsection{Optimization target}
Most networks monitor performance and provide cell-level \glspl{kpi} periodically (e.g., every 15 minutes).
Examples of \gls{ue}-specific uplink \glspl{kpi} include \gls{sinr}, \gls{bler}, bitrate, etc.
Our iterative Bayesian optimization (BO) approach collects these samples from all network cells and processes them in a \gls{cson} server.
Then, based on the data collected, it proposes a new \gls{olpc} configuration for each and all network cells.

Let $\bm{x}=(\alpha, P_0)$ be an \gls{olpc} configuration containing the values of the $\alpha$ and $P_0$ parameters for each network cell.
Then, let $\gamma_{c,u}(\bm{x})$ denote the \gls{pusch} \gls{sinr} of \gls{ue} $u$ in cell $c$ when the network uses configuration $\bm{x}$.
Let $B_{RB}$ denote the bandwidth of one \gls{prb} and let $N_{c,u}$ be the number of uplink \glspl{prb} scheduled for \gls{ue} $u$ in cell $c$.
The bitrate of \gls{ue} $u$ in cell $c$ can either be obtained from network traces or approximated by the Shannon formula:
\begin{equation}
R_{c,u}(\bm{x})=N_{c,u}\cdot B_{RB} \cdot \log_2(1+\gamma_{c,u}(\bm{x})).
\end{equation}

To capture the trade-off between cell-edge and cell-center \gls{ue} performance, we apply an alpha-fairness function (see \cite{Mo2000}) to the collected \gls{kpi} samples.
Alpha-fairness can be applied to any uplink \gls{kpi} (e.g., \gls{sinr}, bitrate, etc).
For example, when applied on bitrate samples, the alpha-fairness metric is
\begin{align}
\Omega^{(r)}(R_{c,u}(\bm{x}))=
\begin{cases}
		10\cdot \log_{10}(R_{c,u}(\bm{x})),& r=1\\
		\frac{R_{c,u}(\bm{x})^{1-r}}{1-r},& r\neq 1\\
\end{cases}\label{eq:fairness}
\end{align}
where $r\geq 0$ is the fairness parameter.

The objective of \gls{olpc} performance optimization is the maximization of a network-wide utility, which we define as the expected alpha-fairness metric perceived by a \gls{ue} with respect to a certain \gls{ue} distribution $\rho$:
\begin{equation}\label{eq:utility}
f(\bm{x}) = \mathbb{E}_{\rho} [ \Omega^{(r)}(R_{c,u}(\bm{x})) ]
\end{equation}
which in practice is approximated via a sample average. 
The optimization problem can then be formulated as:
\begin{equation}
\bm{x}^* = \argmax_x f(\bm{x}).
\end{equation}
It is worth noting that different levels of the fairness parameter $r$ yield optimization problems that are equivalent to optimizing the different Pythagorean means:
\begin{align}
\bm{x}^* = \argmax_x f(\bm{x})=
\begin{cases}
		\argmax_x AM(R_{c,u}(\bm{x})),& r=0\\
		\argmax_x GM(R_{c,u}(\bm{x})),& r=1\\
		\argmax_x HM(R_{c,u}(\bm{x})),& r=2
\end{cases}\label{eq:opt_fairness}
\end{align}
where $AM$, $GM$, and $HM$ are the arithmetic, geometric, and harmonic means, respectively. Moreover, when $r\rightarrow \infty$ the configuration $\bm{x}^*$ tends to the max-min fair solution, for which any increase of the selected KPI for a user causes a KPI drop for a worse-off user.
The $AM$ sensitivity to extreme values is well known, which is why a low level of fairness ($r=0$) maximizes uplink performance for the best-off \glspl{ue} at the expense of \glspl{ue} with lower bitrates.
This sensitivity decreases with the fairness level $r$.
Consequently, \glspl{ue} with low bitrates are the ones that can profit the most from an optimization process with higher fairness.

We remark that, in practice, the \glspl{kpi} measured by commercial networks are not stationary processes.
Traffic demand and network usage often show seasonal patterns, at the time scale of day/week/year.
Therefore, the samples collected by iterative techniques may age quickly and prevent convergence.
This problem can be circumvented via dynamic \gls{rrm} solutions that implicitly learn the seasonality of the traffic evolution. A possible simple solution is to capture the \glspl{kpi} dependence on a heuristically defined internal network state (e.g., hour of day, traffic density, etc.) and train separate models for each such state.
\begin{figure*}[!t]
\centering
\subfloat[\Gls{bogp} convergence on a scenario with $N_c=21$ cells and $2100$ \glspl{ue}.]{\includegraphics[width=0.5 \linewidth]{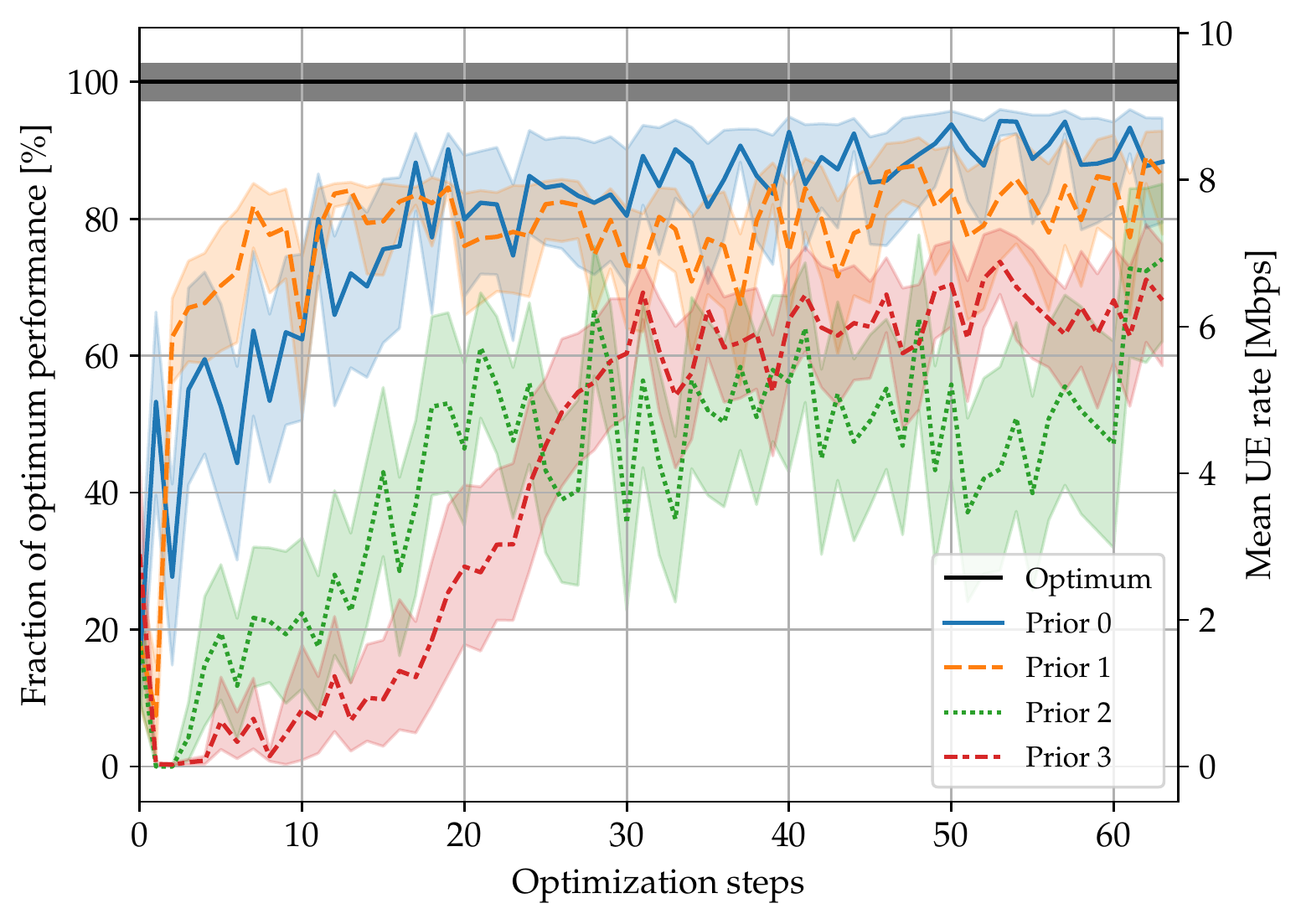}\label{fig:priors_convergence}}
\hfil
\subfloat[Prior mean functions $m(.)$.]{\includegraphics[width=0.5 \linewidth]{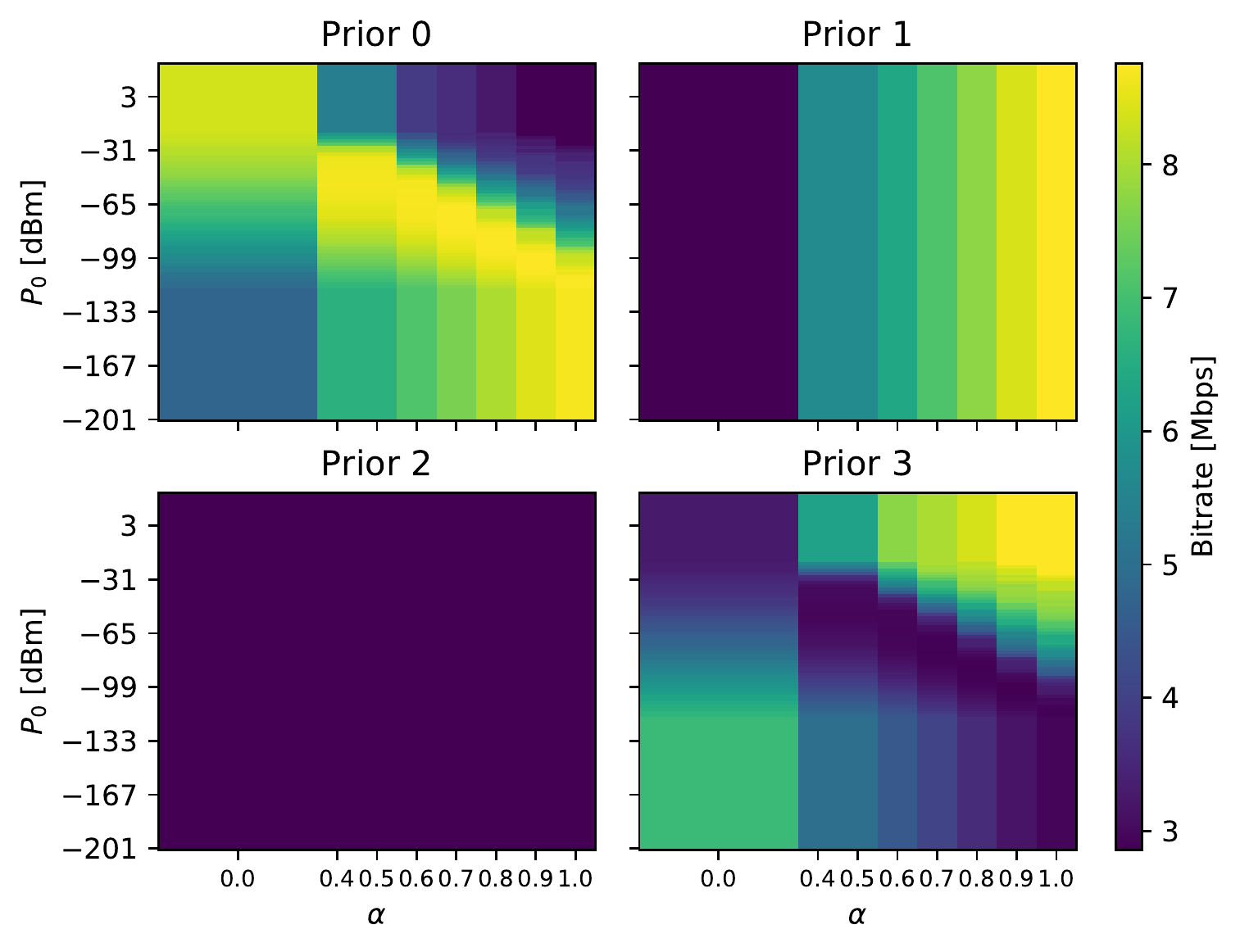}\label{fig:priors_mean}}
\caption{\small \textbf{Impact of prior: Worst-case analysis.} Examples of different priors obtained with a fairness level of $r=0$. Prior 0 has been built with 912 samples from a network with $N_c=3$ cells and 90 UE locations, whereas Prior 1 was built with only 10 samples. Prior 2 has zero mean. Prior 3 is the mirror of Prior 0. All priors share the same covariance matrix, which is a Mat\'ern kernel function $K^{M}_\theta$ as defined in \eqref{eq:matern}.}
\label{fig:priors}
\end{figure*}

%\begin{figure*}[!t]
%	\centering
%	\subfloat[\gls{olpc} configurations evaluated by the optimization algorithm.]{\includegraphics[width=3.5in]{trajectory.pdf}\label{fig:trajectory}}
%	\hfil
%	\subfloat[Convergence time comparison. The solid line depicts the mean performance and the shadowed region the 95\% confidence interval.]{\includegraphics[width=2.5in]{convergence.pdf}\label{fig:convergence}}
%	\caption{\textbf{BOGP convergence.} Evaluation with $K=4$ active \glspl{ue} per snapshot and a fairness level of $r=1$ on the network architecture of Fig.~\ref{fig:network}. Results for \gls{bogp} with and without a prior are shown. A \gls{cd} algorithm with \gls{gss} on each coordinate is also shown for reference.}
%	\label{fig:performance}
%\end{figure*}

\subsection{System architecture \& algorithm}

We propose to deploy a centralised self-organizing network (\Gls{cson}) server that collects cell-level uplink \glspl{kpi} from the cells to be optimized and proposes a new \gls{olpc} configuration in return.
The time duration during which \glspl{kpi} are collected is called a sampling period and for scalability reasons, all cells are provisioned with the same \gls{olpc} configuration.
We also experimented with grouping the cells into clusters and assigning different \gls{olpc} configurations to different clusters.
However, this increased the size of the problem and convergence time by several orders of magnitude, in exchange for only marginal performance gains.

Our iterative \gls{bo} algorithm would run in the \gls{cson} server.
At each sampling period $n$, it summarizes all the \gls{kpi} samples collected during that period into the sample average $\widetilde f(\bm{x}(n))$  \eqref{eq:utility}.
This way, a database of configuration to performance mappings $\mathcal{D}=\{\bm{x}(n), \widetilde f(\bm{x}(n))\}_{\forall n}$ is progressively built and used to train a \gls{gp} model of the system performance function.
The configuration $\bm{x}(n+1)$ to try at the beginning of a new period $n+1$ is chosen such that an acquisition function is maximized as described in great detail in Section~\ref{sec:BOdetails}.

\subsection{Simulation setup \& Results} \label{sec:results}

We tested our algorithm in a simulated cellular network with $N_{sites}=7$ sites (see Fig.~\ref{fig:network}), $N_{sectors}=3$ cells per site, and $N_b=32$ beams per cell.
There are $N_c=N_{sites}\cdot N_{sectors}=21$ cells in total.
The carrier frequency was 3.5 GHz and the system bandwidth contained 100 physical resource blocks (\glspl{prb}) of bandwidth 180 kHz each. The maximum transmit power per \gls{ue} was $P_{CMAX}=23$ dBm and we used an \gls{umi} 3D path loss model based on Table 7.4.1-1 of 3GPP TR 38.901. In our network, 80\% of \glspl{ue} were indoor and 20\% of \glspl{ue} were outdoor and the \gls{los} probability was computed as described in section 7.4.2 of 3GPP TR 38.901.
We simulated a total of 2100 \glspl{ue}, with an average of 100 \glspl{ue} per cell, although not all \glspl{ue} were connected to their serving cells all the time.
A wrap-around radio distance model has been used to reduce interference bias for \glspl{ue} at the edge of the network.
At each snapshot, a set of $K\in[2,\dots,16]$ \glspl{ue} per cell was sampled uniformly at random from the set of all \glspl{ue} served by that cell.
These parameter values have been chosen to simulate a network that is dense enough (both in terms of the number of cells and \glspl{ue}) to be interference limited in the uplink. It is in such networks where the difference between poor and good power control can be observed more clearly. The power and spectral values resemble those from commercial systems.

At each sampling period, the base stations (\glspl{bs}) were provisioned with a new \gls{olpc} configuration chosen by the \gls{bogp} algorithm. Then, $S=16$ snapshots were simulated with that configuration. $S$ reflects the number of \gls{pm} counters a C-SON server may be able to collect during a single day.
This yields a total of $S \cdot N_c \cdot K$ uplink bitrate samples per sampling period.
During each snapshot, the cells try to schedule \glspl{prb} equally to each of their served \gls{ue}, while respecting their power limits (i.e., cell-edge \glspl{ue} may not have enough uplink power to use as many \glspl{prb} as cell-center \glspl{ue}).

First, we used exhaustive search to explore the target function defined in \eqref{eq:utility} in the full $(\alpha, P_0)$ solution space.
This produced the system performance function $f(\bm{x})$ illustrated in Fig.~\ref{fig:target} which shows, among other traits, that $f$ is not unimodal.
The locations of the optimum \gls{olpc} configurations for various fairness levels $r$ are also illustrated.
Despite this being an idealized simulated scenario, i) the multi-modality of $f(\bm{x})$ and ii) the fact that only a noisy version $\widetilde f$ of it is observed already pose a challenge for traditional direct-search derivative-free algorithms (e.g., binary search, golden section search search).

Fig.~\ref{fig:performance} compares the performance of \gls{bogp} for fairness $r=1$ against that of the coordinate-descent Golden Section Search (\gls{cd}-\gls{gss}) optimization algorithm \cite{conn2009introduction}, a well known direct-search derivative-free method that we use as a baseline.
\gls{cd}-\gls{gss} optimizes over the $\alpha$ and $P_0$ coordinates in successive fashion, using \gls{gss} on each dimension.
Within one coordinate pass and in the absence of observation noise, \gls{cd}-\gls{gss} is capable to attain a local maximum with respect to the specific coordinate. However, after a coordinate change the performance drops suddenly, since a line search is performed from scratch with respect to the new coordinate. Indeed, we argue that direct-search methods are especially useful for \emph{offline} derivative-free optimization, since they do not fulfill the properties P2-P3 outlined in Section \ref{sec:benefitsBO}. 

In contrast, once BOGP has attained a near-optimal performance, it automatically reduces considerably the search region and avoids perilous and (almost) pointless further exploration. Moreover, BOGP achieves, on average, 90\% of the optimum performance in barely 20 evaluations (see Fig.~\ref{fig:convergence}).
To achieve this, we used a covariance Mat\'ern kernel function $K^{M}_\theta$ defined as in \eqref{eq:matern}. We simply initialized the hyperparameters $\ell$ so as to homogenize the scale of the two variables of $\xb=[\alpha,P_0]$, i.e., $\ell_\alpha/\ell_{P_0}=226$.
Fine tuning of the covariance kernel hyperparameter $\ell$ could possibly bring convergence closer to the optimum performance. Hyperparameter tuning can be done in simulation and prior to the actual optimization on a commercial network. However, fine-tuning $\ell$ on the live network would be hard to justify to an operator for a mere gain increase of 90\% to 100\%.
In this experiment, we employed the expected improvement (EI) acquisition function in its most risk-sensitive variant, i.e., with $\xi=0$.

Fig.~\ref{fig:convergence} shows the \gls{bogp} performance when the \gls{gp} is initialized with a prior obtained from a different and smaller network with only $N_c=3$ cells and 90 \gls{ue} locations for each $(\alpha,P_0)$ pair. 
Despite the major architectural differences and traffic requirements between the two networks, Fig.~\ref{fig:convergence} shows that prior reuse helps to improve performance in the initial steps and to quickly steer the optimization process towards a good $(\alpha,P_0)$ region.
The solution regions explored by the \gls{bogp} are illustrated in \cref{fig:trajectory}.

No \gls{mno} would ever deploy an algorithm that drives uplink performance to 0, even if high gains can be obtained later on.
We therefore see the reuse of priors as one of the main advantages of the \gls{bogp} algorithm. In a real network setup, one could imagine that clusters of cells would share their priors to speed up the optimization of other clusters. 

Finally, we illustrate in Fig.~\ref{fig:priors} the impact of using different, and possibly off-the-mark, priors when starting the optimization process. A fairness parameter $r=0$ is used here.
As expected, priors with a mean close to that of the target function help to increase the performance very early in the optimization process (see Prior 0 in Fig.~\ref{fig:priors}, obtained in the same 3 cell setup as above).
As mentioned above, such priors can be obtained by collecting data from previous network deployments.
Remarkably, even simplified priors---such as Prior 1 which was obtained by interpolating Prior 0 on few samples---help to accelerate convergence, suggesting that the prior does not need to match accurately the actual performance function $f$. Rather, the prior should hint at the best regions that BO has to concentrate on.

As a worst-case analysis, priors that differ substantially from the target function (e.g., Prior 3 in Fig.~\ref{fig:priors}, being the opposite of Prior 0) slow down convergence. In this case, BO is taught to focus on a low performance $(\alpha,P_0)$ region, which BO can escape only by random exploration, i.e., by setting the parameter $\xi>0$ and sufficiently high. Eventually, the algorithm is capable of learning the right trend and it overcomes the initial bad estimates, although the slow convergence may be unacceptable for most \glspl{mno}. Yet, we argue that this would constitute an extreme and unrealistic case. In fact, if the controller suspects that the available prior information is unreliable, then it should simply settle for a constant prior mean $m$ (see Fig.~\ref{fig:performance} and Prior 2 in Fig.~\ref{fig:priors}).\\

\begin{figure}
	\centering
	\includegraphics[width=\linewidth]{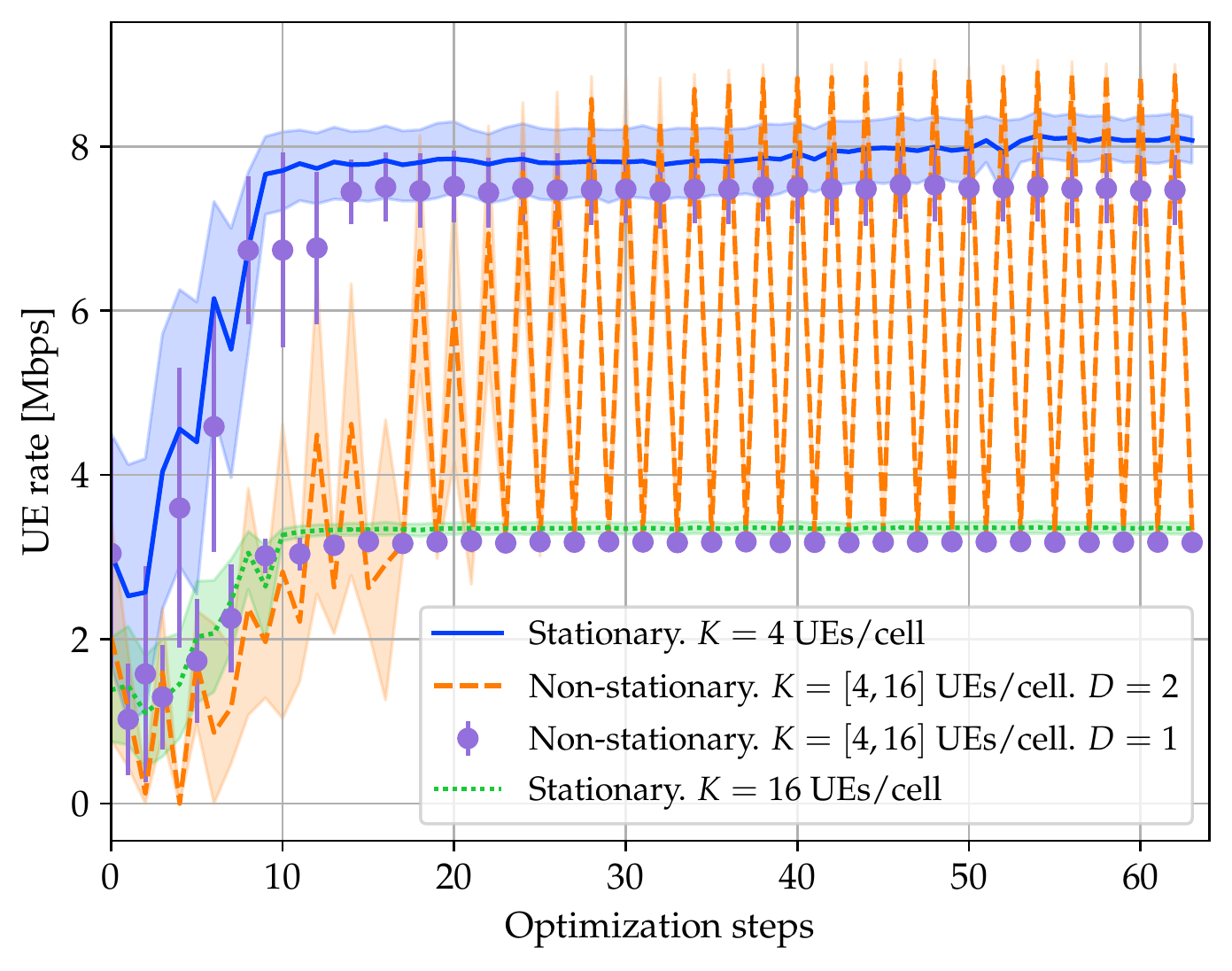}
	\caption{\small \textbf{Convergence in a load-varying scenario.} In the stationary case, the number $K$ of active \glspl{ue} per cell and TTI remains constant on all optimization steps. In the non-stationary case, it changes at every step (e.g., peak vs off-peak hours). The case $D=1$ coincides with the classic, non-dynamic BOGP algorithm.}
	\label{fig:dbo}
\end{figure}

We also evaluated the BOGP performance in a dynamic setting where the average cell load varies over time. Using the notation of Section \ref{sec:timevarying}, the network \emph{state} $\theta(t)$ is represented by the cell load at time $t$. The state $\theta$ varies according to a periodic pattern of period 2, namely at each sampling time $t_n$ the average cell load alternates between $K=4$ and $K=16$ UEs per TTI. We still used a Mat\'ern kernel for computing the covariance between different RRM configurations, while we employed the Exp-Sine-Squared kernel \eqref{eq:ESS} as time kernel. 
We let BOGP optimize autonomously and in an online fashion the length scale $\ell_{ESS}$ of the time kernel---regulating the covariance between off-period samples---as described in Section \ref{sec:hyparam_tuning}. In Figure \ref{fig:dbo} we illustrate the convergence of the classic \emph{static} \gls{bogp} (\emph{non-stationary}, time kernel period $D=1$, in the legend) and the \emph{dynamic} and periodic BOGP ($D=2$), where each point is averaged over 16 independent runs with random initialization points and a constant prior mean. 
As a benchmark, we also show the case where two BOGPs are run in parallel (\emph{stationary} case, in Figure \ref{fig:dbo}), one in each state with $K=4,16$, respectively. Clearly, the latter scenario is implementable only if the state evolution is known in advance. We set the UCB parameter $\beta_n=1$ for each iteration $n$ (cfr. Eq. \eqref{eq:UCB}). Classic BO is agnostic to the time evolution and finds a \emph{single} RRM configuration with good performance in \emph{both} states, while dynamic BO manages to distinguish between the two scenarios and adapts its RRM configuration accordingly.
Letting dynamic BO ($D=2$) optimize the time length scale $\ell_{ESS}$ produces two effects: i) dynamic BO suffers from cold start, while it figures out the best value for $\ell_{ESS}$; ii) from iteration 25 onward, the dynamic BO outperforms most of the other approaches since it exploits the correlation between the two scenarios to learn a better configuration for the less loaded scenario ($K=4$). We think that the gains will be even more accentuated in network deployments with larger load variations (e.g., shopping malls). 
We remark that in practical \gls{rrm} settings, network traffic shows seasonal patterns whose periodicity can be inferred with good confidence via historical data and is typically circadian.
% \gls{bogp} manages to learn the correct periodicity within 20 iterations and improves towards the optimal objective values for both cell load states. 

\section{Conclusions and future directions}
\label{sec:conclusions}

We provided an accessible introduction to Bayesian optimization with Gaussian Processes (\gls{bogp}). Moreover we showed under which conditions \gls{bogp} can serve the purpose of tuning RRM parameters in online fashion, by minimizing the risk of incurring performance drops. We instantiated BOGP on a realistic uplink power control setting, where the goal is to optimize two network-wide parameters describing the target received power at the base station and the path loss compensation factor.

We believe that \gls{bogp} is a powerful and versatile tool that has not yet received the attention that it deserves from our community. We think that there exist at least three research directions that can pave the way for the adoption of BOGP over a larger set of RRM use cases. The first one is the incorporation of hard constraints---typically, on the user Quality of Service---that are also modeled as black-box functions that can only be learned via noisy sampling. We deem the recent work in \cite{constrainedBOnoisy} as an excellent starting point.
Another relevant direction, extending our dynamic BO analysis, consists in directly embedding the ``state'' of the network in the GP to efficiently transfer the learning across different states, as first studied in \cite{krause2011}. Last but not least, if sufficient historical data is available, a \emph{parametric} model could be built (in contrast to GPs, which are \emph{non-parametric}) that is well representative of the objective function. Different Bayesian techniques such as Thompson sampling can be used to find the next point to sample, as reported, e.g., in \cite{Shahriari2016}.

\section*{Acknowledgments}
We would like to thank our colleagues Véronique Capdevielle, Richa Gupta, and Suresh Kalyanasundaram for insightful discussions on the topic of \gls{olpc} and the pros \& cons of \gls{bogp} for RRM.


\end{document}